\documentclass[11pt,a4paper]{article}
\pdfoutput=1

\usepackage{amsmath}
\usepackage{amsfonts}
\usepackage{amsthm}
\usepackage{amssymb}
\usepackage{amscd}
\usepackage[british]{babel}
\usepackage{graphicx}
\usepackage{psfrag}
\usepackage{epsfig}
\usepackage{rotating}
\usepackage{times}
\usepackage{color}
\usepackage{float}
\usepackage{ulem}

\DeclareMathOperator{\arcsinh}{Arsh}

\theoremstyle{plain}

\newtheorem{remark}{Remark}[section]

\newcommand{\boxend}{\flushright{$\Box$}}

\newcommand{\N}{{\mathbb N}}               
\newcommand{\R}{{\mathbb R}}               

\renewcommand{\tilde}{\widetilde}

\begin{document}

\title{The matter-ekpyrotic bounce scenario in Loop Quantum Cosmology}

\author{
Jaume Haro$^{a,}$\footnote{E-mail: jaime.haro@upc.edu},
 Jaume Amor\'os$^{a,}$\footnote{E-mail: jaume.amoros@upc.edu}
and 
Llibert Arest\'e Sal\'o$^{a,}$ \footnote{E-mail: llibert.areste@estudiant.upc.edu} 
}

\maketitle

\begin{center}
{\small

$^a$Departament de Matem\`atica Aplicada, Universitat
Polit\`ecnica de Catalunya \\ Diagonal 647, 08028 Barcelona, Spain \\

}
\end{center}

\thispagestyle{empty}

\begin{abstract}
We will perform a detailed study of the matter-ekpyrotic bouncing scenario in Loop Quantum Cosmology using the methods of the dynamical systems theory. We will show that 
when the background is driven by a single scalar field,
 at very late times, in the contracting phase, all orbits depict  a matter dominated Universe, which evolves to an ekpyrotic phase. 
 After the bounce the Universe enters in the expanding phase, where the orbits leave the ekpyrotic regime going to a kination (also named deflationary) regime.
 Moreover, this scenario supports the  production of heavy massive particles conformally coupled with gravity, which reheats the universe at temperatures compatible with
 the nucleosynthesis bounds and also the production of massless particles non-conformally coupled with gravity leading  to very high reheating temperatures but  ensuring the
 nucleosynthesis success.
 Dealing with cosmological perturbations, 
 these background dynamics produce a nearly scale invariant power spectrum for the modes that leave the Hubble radius, in the contracting phase, 
 when the Universe is quasi-matter dominated, whose spectral index and corresponding running 
  is compatible with the recent experimental data obtained by PLANCK's team.
\end{abstract}

\vspace{0.5cm}

{\bf Pacs numbers:}

\section{Introduction}

Bouncing cosmologies and more specifically the matter bounce scenario (see \cite{brandenberger} for a review of this kind of cosmologies) is the most promising alternative to the 
inflationary 
paradigm. There are many ways to obtain cosmologies without the Big Bang singularity: for example  violating the null energy condition in General Relativity by 
incorporating  new forms of matter such as phantom \cite{a} or quintom  fields \cite{b}, Galileons \cite{c}
or phantom condensates \cite{d}, or by adding terms to Einstein-Hilbert action \cite{roshan}, but the simplest one is to go beyond General Relativity and consider holonomy corrected 
Loop Quantum Cosmology (LQC), 
where a Big Bounce replaces the Big Bang singularity \cite{singh}. In fact, other future singularities such as Type I (Big Rip) and Type III (Big Freeze) are also forbbiden in 
holonomy corrected LQC \cite{Haro1}.

\

On the other hand,
it is well known that a matter domination period in the contracting phase is dual to the de Sitter regime in the expanding phase \cite{wands}, which provides a flat power spectrum
of cosmological perturbations
in the matter bounce scenario. Moreover, an abrupt ekpyrotic phase transition is needed, in the contracting phase, in order to solve the famous Belinsky-Khalatnikov-Lifshitz (BKL) instability:
 the effective energy density of primordial anisotropy scales as $a^{-6}$ in the contracting phase \cite{Belinsky:1970ew} and, more important, to produce enough particles 
 which will be responsible for thermalizing the universe
 in the expanding phase
 \cite{haro-elizalde,he}. This new scenario named matter-ekpyrotic scenario was introduced in \cite{cai3}, being developed within the two-field model in \cite{cai4}, while in \cite{cai5} there is a numerical discussion on the primordial anisotropy issue (see \cite{cai6} for a review). It was treated in the context of Loop Quantum Cosmology in \cite{wilson2}, showing 
 that it could be compatible with 
 the new experimental data \cite{Ade} provided by PLANCK's team (see also \cite{ewing}).

\

The main goal of the present work is to study mathematically, from the viewpoint of the dynamical systems theory, the matter-ekpyrotic scenario in the context of holonomy corrected LQC.
More precisely, we study its background dynamics and 
the evolution of cosmological scalar perturbations in this scenario, showing that it leads to a nearly flat power spectrum of perturbations, 
whose spectral index and running match correctly with the recent experimental data. Moreover, we will also point out that the phase transition, in the contracting epoch, from matter domination to the ekpyrotic phase, is the responsible for the production of enough particles whose energy density, in the expanding phase, will be dominant, leading to a reheating temperature compatible with the nucleosynthesis bounds.

\

The work is organized as follows: 
The dynamical study of the matter-ekpyrotic scenario in General Relativity is performed in section II.
In section III we study the dynamics of the background,  demonstrating that in Loop Quantum Cosmology the matter-ekpyrotic scenario has two different regimes: at very early times,  the universe in the contracting phase is  matter dominated and evolves to an ekpyrotic regime, and after the bounce, we show that the universe leaves the ekpyrotic phase and enters in a kination (or deflationary) regime \cite{spokoiny}. Section IV is devoted to the calculation of the power spectrum of scalar perturbations in this scenario, showing that the power spectrum is proportional to the square of the
value of the Hubble parameter at the transition time. The mechanism to reheat the universe is studied in section V, where we show that the phase transition from the matter to the ekpyrotic regime is essential to produce enough particles to reheat the universe. In last section, we build a simple potential in this scenario that leads to theoretical values of the spectral index and its running that fit well with the last observational data provided by PLANCK'S team.
 
 \vspace{1cm}
 
The units used throughout  the paper are $\hbar=c=8\pi G=1$  and,  thus, the reduced Planck's mass is  $M_{pl}=\frac{1}{\sqrt{8\pi G}}=1$.

\section{Matter-ekpyrotic scenario in General Relativity: The background}
In this section,
we consider the flat Friedmann-Lema{\^\i}tre-Robertson-Walker (FLRW) metric in the context of General Relativity. Then, the Friedmann, conservation and Raychauduri equations are 
respectively given by
\begin{eqnarray}
H^2=\frac{\rho}{3}, \quad \dot\rho=-3H(\rho+P),\quad \dot{H}=-\frac{1}{2}(\rho+P),
\end{eqnarray}
where $H$ is the Hubble parameter, $\rho$ the energy density and $P$ the pressure.

To achieve the matter-ekpyrotic scenario we will deal with a single scalar field, namely $\varphi$. Then, the energy density and pressure are
\begin{eqnarray}
\rho=\frac{\dot{\varphi}^2}{2}+V(\varphi),\quad P=\frac{\dot{\varphi}^2}{2}-V(\varphi),
\end{eqnarray}
being $V(\varphi)$ the potential.

Our first goal is
 to mimic, with a single scalar field $\varphi$, a fluid with Equation of State (EoS) $P=w_0\rho$, i.e., we want to obtain a potential $V(\varphi)$ which provides a solution of the
conservation equation that
leads to the same background as a barotropic fluid with EoS $P=w_0\rho$, where $w_0>-1$.

The first step is to write the EoS $P=w_0\rho$ as follows {
\begin{eqnarray}w_0=\frac{P}{\rho}=\frac{\frac{\dot{\varphi}}{2}-V(\varphi)}{ \frac{\dot{\varphi}}{2}+V(\varphi) }   \Longrightarrow V(\varphi)=\frac{1-w_0}{1+w_0}\frac{\dot{\varphi}^2}{2},
\end{eqnarray}}
which means that for $-1<w_0<1$ the potential must be positive, negative for $w_0>1$, and vanishes for $w_0=1$.

On the other hand, for a fluid with  EoS $P=w_0\rho$, the Friedmann and conservation equations of General Relativity
could be easily solved,  giving as a result the background
\begin{eqnarray}a(t)\propto |t|^{\frac{2}{3(1+w_0)}}\Longrightarrow H(t)=\frac{2}{3(1+w_0)t}.\end{eqnarray}

Secondly, using the Raychauduri equation  $\dot{H}=-\frac{\dot{\varphi}^2}{2}$ one has $\dot{\varphi}=\pm\sqrt{-2\dot{H}}$, and
if we only consider increasing functions, i.e., $\dot{\varphi}=\frac{2}{\sqrt{3(1+w_0)}}\frac{1}{|t|}$
we will obtain
\begin{eqnarray}\label{5}\varphi_{\pm}(t)=\pm \frac{1}{\sqrt{3(1+w_0)}}\ln\left(t/t_0\right)^2,
\end{eqnarray}
where $-$ (resp. $+$) refers to the contracting  (resp. expanding) phase. In fact,  $\varphi_{-}$ is defined for $t\in(-\infty,0)$, and $\varphi_{+}$
in $t\in(0, \infty)$.

Finally, to reconstruct the potential we use the relation $P=w_0\rho$ and the Raychaudhuri equation, to obtain
\begin{eqnarray}
 V=\frac{w_0-1}{1+w_0}\dot{H}\Longrightarrow V=\frac{2(1-w_0)}{3(1+w_0)^2}\frac{1}{t^2}.
\end{eqnarray}

Using \eqref{5}, a simple calculation leads to
\begin{eqnarray} V_{\pm}(\varphi)=\frac{2(1-w_0)}{3(1+w_0)^2}\frac{1}{t_0^2}e^{\mp \sqrt{3(1+w_0)}\varphi},
\end{eqnarray}
and, by writing {
\begin{eqnarray}
V_{\pm}(\varphi)=V_0e^{\mp \sqrt{3(1+w_0)}\varphi},
\end{eqnarray}} the analytical solution \eqref{5} becomes
\begin{eqnarray}\label{8} \varphi_{\pm}(t)=\pm \frac{1}{\sqrt{3(1+w_0)}}\ln\left({\frac{3V_0(1+w_0)^2}{2(1-w_0)}}t^2\right).\end{eqnarray}

{ We note that this equation is mathematically singular for $w_0=1$, which should be treated separately. In this case, since $V=0$ one cannot use the parameter $V_0$ but is only able express the scalar fields through the parameter $t_0$, as shown in \eqref{5}. On the other hand, with regards to the trivial case of $w_0=-1$, it corresponds to a constant $H$, a constant scalar field and a constant potential $V=V_0$. }

\subsection{Dynamical analysis}

Once we have obtained the potential 
for a single scalar field, the dynamical equations (Friedmann and conservation) are
\begin{eqnarray}
H^2=\frac{1}{3}\left(\frac{\dot{\varphi}^2}{2}+V\right);\quad \ddot{\varphi}+3H_{\pm}\dot{\varphi}+V_{\varphi}=0,\end{eqnarray}
where $H_{\pm}=\pm \frac{1}{\sqrt{3}}\sqrt{\frac{\dot{\varphi}^2}{2}+V(\varphi)}$.

The conservation equation is a second order differential equation, meaning that there are infinitely many solutions that depict different universes. For example, if one deals in 
the expanding
phase, i.e. we take $H_+$, then for the potential $V=V_0 e^{-\sqrt{3(1+w_0)}\varphi}$, the analytical solution 
$\varphi_{+}(t)= \frac{1}{\sqrt{3(1+w_0)}}\ln\left({\frac{3V_0(1+w_0)^2}{2(1-w_0)}}t^2\right)$ with $t>0$, defines a Universe with EoS $P=w_0\rho$ all the time. However, the other solutions
do not define a Universe with this EoS all the time.

For this reason, what we want to study is the behaviour of the analytical solution \eqref{8}, i.e., we want to know whether the orbit 
in the plane $(\varphi,\dot\varphi)$
defined by the solution (\ref{8}) will be an attractor or a repeller.

To do that, first of all we study the dynamics in the contracting phase. Performing the change of variable \cite{hae}
\begin{eqnarray}\varphi=-\frac{2}{\sqrt{3(1+w_0)}}\ln \psi,
\end{eqnarray}
the conservation equation will become
\begin{eqnarray}
\frac{\ddot\psi}{\psi}-\frac{\dot\psi^2}{\psi^2}-\sqrt{3}\sqrt{\frac{2}{3(1+w_0)}\frac{\dot\psi^2}{\psi^2}+\frac{V_0}{\psi^2}}\frac{\dot\psi}{\psi}-\frac{3(1+w_0)}{2}\frac{V_0}{\psi^2}=0,
\end{eqnarray}
that could be written as a first order one-dimensional differential equation:
\begin{eqnarray} \frac{d\dot\psi}{d\varphi}=F_{-}(\dot\psi),
\end{eqnarray}
where
\begin{eqnarray}
F_{-}(\dot\psi)=-\frac{3}{2}\sqrt{1+w_0}\left[\sqrt{\frac{2{\dot\psi^2}}{3(1+w_0)}+{V_0}}+\frac{\sqrt{3}(1+w_0)}{2\dot\psi}
\left(\frac{2{\dot\psi^2}}{3(1+w_0)}+{V_0}\right)
\right].
\end{eqnarray}

To perform the dynamical study we have to differentiate between $3$  cases:
\begin{enumerate}
 \item  $w_0=1$.
 In this case the differential equation becomes:
 \begin{eqnarray}\frac{d\dot\psi}{d\varphi}=-\frac{\sqrt{3}}{\sqrt{2}}\left[|\dot\psi|+\dot\psi\right],\end{eqnarray}
which means that in the half plane $\dot{\varphi}\geq 0\Longleftrightarrow\dot\psi\leq 0$, one has $\frac{d\dot\psi}{d\varphi}=0\Longrightarrow \dot{\psi}=-|C_0|$ and
the general solution is given by
\begin{eqnarray}\psi(t)=-|C_0|t+|C_1| \Longleftrightarrow \varphi(t)=-\sqrt{\frac{2}{3}}\ln (-|C_0|t+|C_1| ).
\end{eqnarray}

{
That is, there are two kind of orbits in the plane $(\varphi,\dot\varphi)$
\begin{eqnarray}\gamma_1(t)=\left(-\sqrt{\frac{2}{3}}\ln (-|C_0|t+ |C_1|),\sqrt{\frac{2}{3}}\frac{1}{-t+|C_2|}\right)\quad \mbox{with} \quad t<|C_2|,
\end{eqnarray} with $|C_2|\equiv \frac{|C_1|}{|C_0|}$,
and
\begin{eqnarray}\gamma_2(t)=(|C_1|,0)\quad \mbox{with} \quad t\in \R.
\end{eqnarray}

}

The first kind corresponds to a stiff fluid because  $\gamma_1$ leads to $H(t)=\frac{1}{3t}$. And the second kind are fixed points that correspond to  $H=0$.

On the other hand, in the plane $\dot\varphi<0$,  it is easier to study directly the conservation equation
\begin{eqnarray}
 \ddot{\varphi}+3H_-\dot{\varphi}=0\Longleftrightarrow \ddot{\varphi}+\sqrt{\frac{3}{2}}\dot{\varphi}^2=0,
\end{eqnarray}
obtaining {
\begin{eqnarray}\gamma_3(t)=\left(\sqrt{\frac{2}{3}}\ln (-|C_0|t+|C_1|),\sqrt{\frac{2}{3}}\frac{1}{-t+|C_2|}\right)\quad \mbox{with} \quad t<|C_2|.
\end{eqnarray}

We point out that orbits $\gamma_1$ and $\gamma_3$ are not stable in the following sense: We consider for example $\gamma_3$ and we make a small perturbation
taking $\bar{|C_2|}\equiv |C_2|+\delta |C_2|$, then the modulus of the  difference between $X(t)\equiv \sqrt{\frac{2}{3}}\frac{1}{-t+|C_2|}$ and 
$\bar{X}(t)\equiv \sqrt{\frac{2}{3}}\frac{1}{-t+\bar{|C_2|}}$, namely $\delta X(t)$, is  equal, at the leading order, to 
$\sqrt{\frac{2}{3}}\frac{\delta |C_2|}{(-t+{|C_2|})^2}$.
Therefore, one can trivially see that $\frac{\delta X(t)}{X(t)}\to\infty$ as $t\to |C_2|$.}
Nevertheless, what is important is that all the orbits in the phase space for $w_0=1$ correspond to a universe which obeys throughout all the evolution of time to the same Equation of state, namely $P=\rho$. 

\begin{figure}[H]
\begin{center}
\includegraphics[scale=0.25]{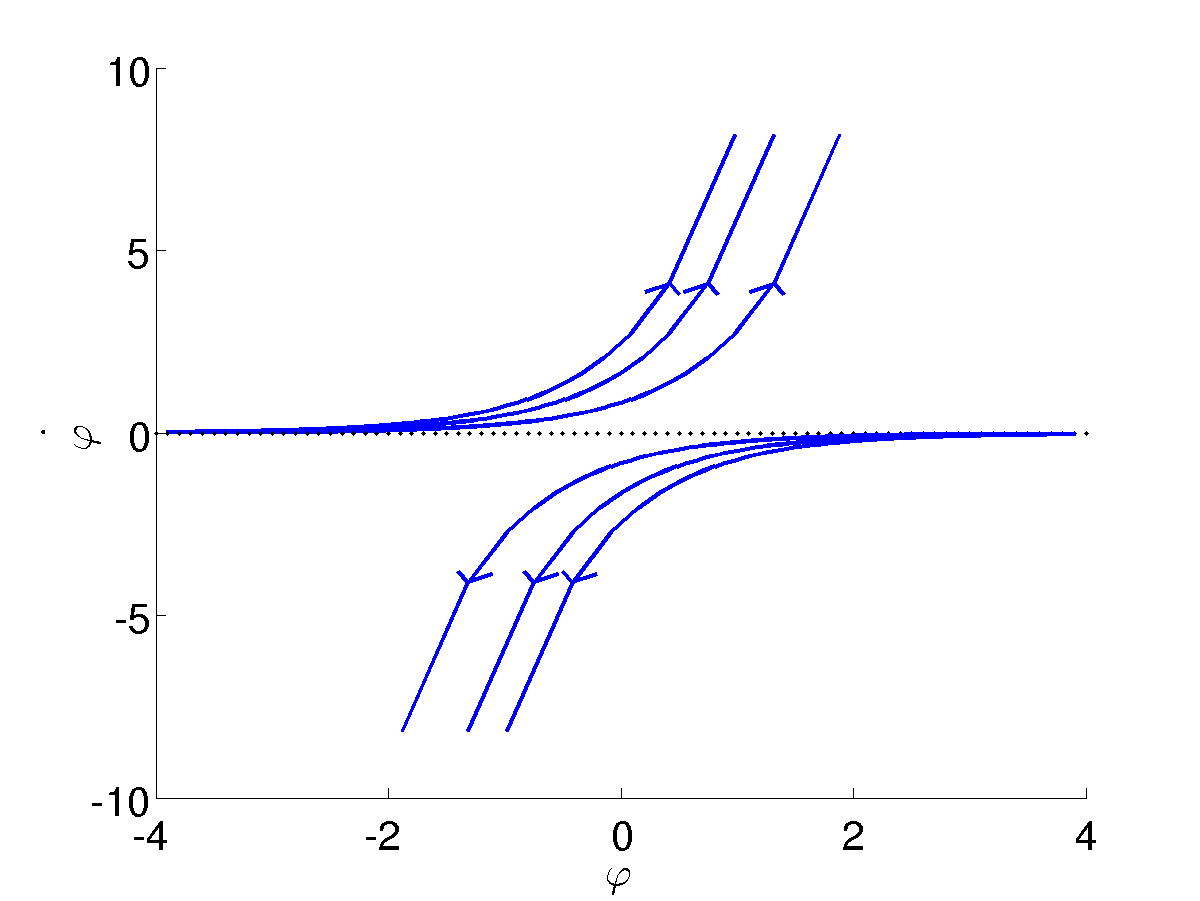}
\includegraphics[scale=0.30]{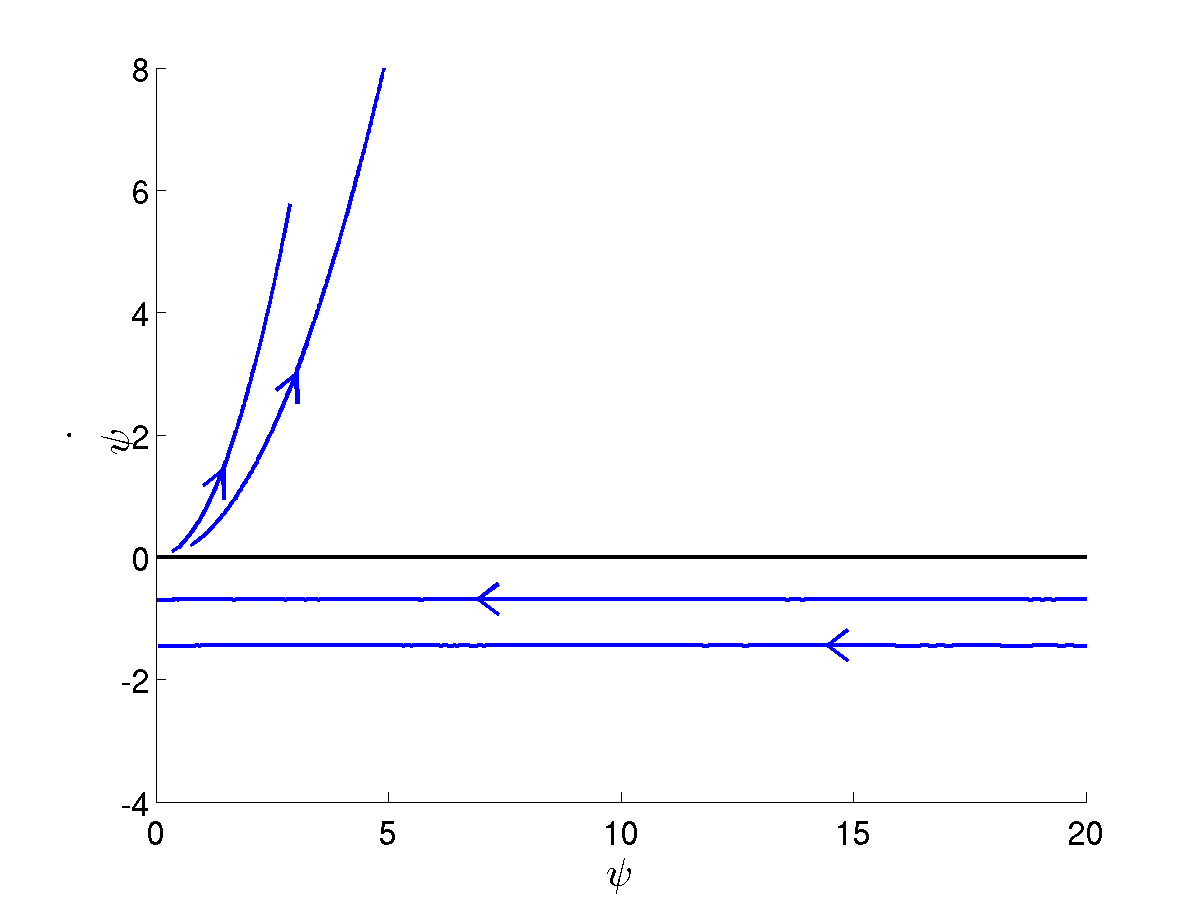}
\end{center}

\caption{Phase portrait in the $(\varphi,\dot{\varphi})$ plane (left) and in the $(\psi,\dot{\psi})$ plane (right) in the case $w_0=1$. The orbits of the families $\gamma_1$ (blue, upper semiplane) and $\gamma_3$ (blue, lower semiplane), separated by the equilibrium points $\gamma_2$ (black dots) fill the plane.}
\label{retratgrw1}
\end{figure}


\item $-1<w_0<1$.
The system has one fixed point, namely
\begin{eqnarray}\dot\psi_1=-(1+w_0)\sqrt{\frac{3V_0}{2(1-w_0)}},
\end{eqnarray}
which is a global repeller because $F_{-}(0^-)=-F_{-}(-\infty)=+\infty$,  and for $\dot{\psi}>0$, $F_{-}$ is negative, which means that in cosmic time the orbits move away from 
$\dot\psi_1$.

This solution corresponds to
\begin{eqnarray}\label{19}\varphi_1(t)=- \frac{1}{\sqrt{3(1+w_0)}}\ln\left({\frac{3V_0(1+w_0)^2}{2(1-w_0)}}t^2\right),
\end{eqnarray}
whose orbit depicts, all the time, in the contracting phase, a Universe with EoS $P=w_0\rho$.

Note that the fact that solution \eqref{19} is a repeller means that all orbits depict, at early times, { a universe with equation of state $P=w_0\rho$, which will be a matter dominated Universe in the particular case when $w_0=0$. Moreover, since time goes forward, the solutions 
separate from \eqref{19}, meaning 
that the backgrounds depicted by them cease to satisfy this equation of state.} The phase portrait is drawn in Figure \ref{retratgrw0}.

\begin{figure}[H]
\begin{center}
\includegraphics[scale=0.30]{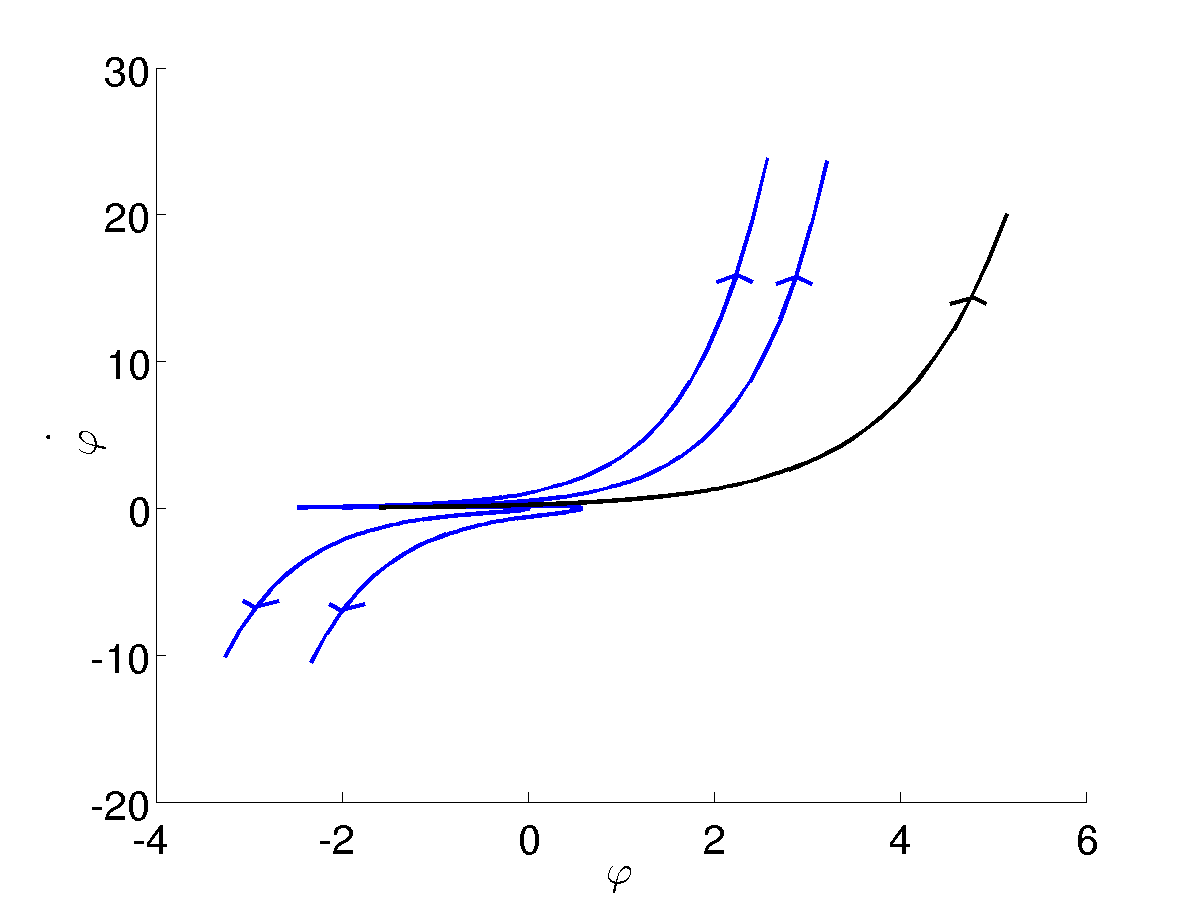}
\includegraphics[scale=0.30]{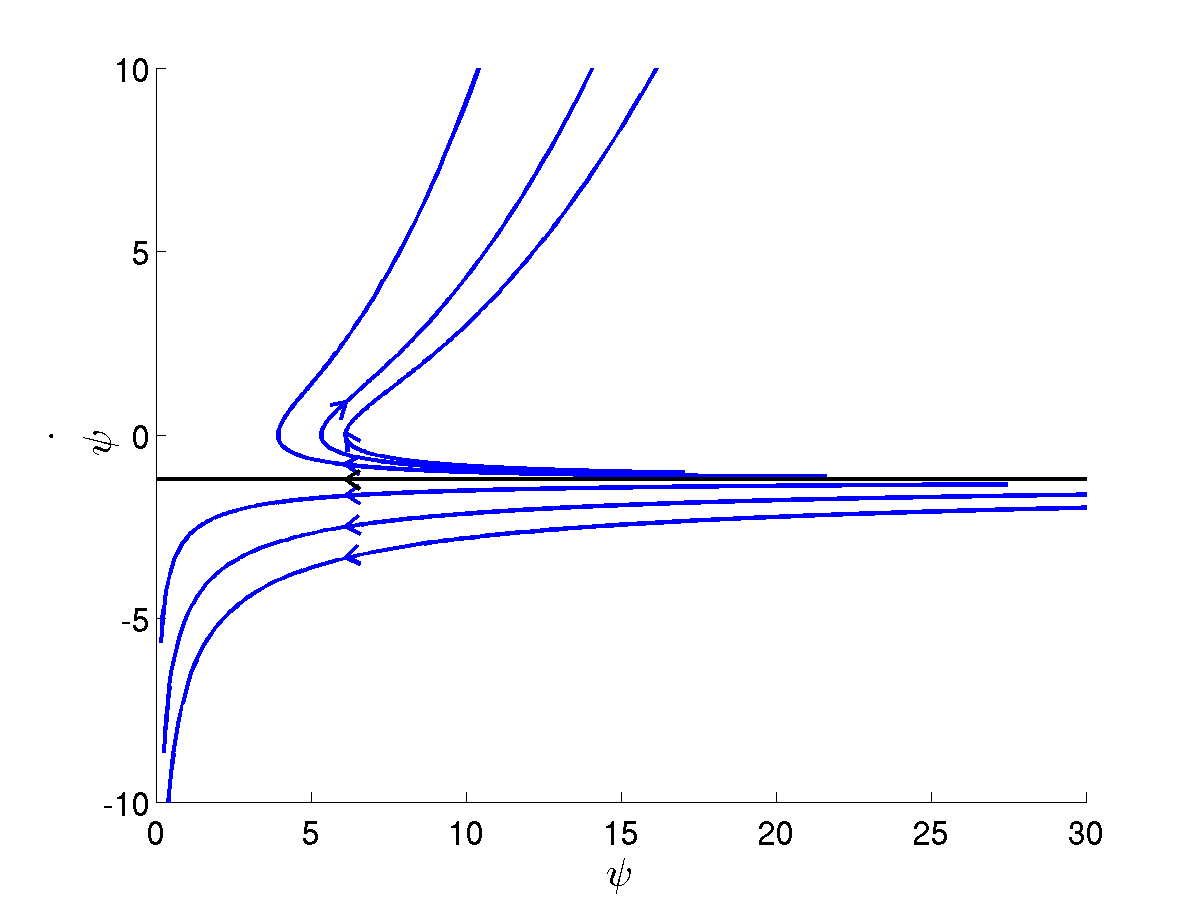}
\end{center}

\caption{Phase portrait in the $(\varphi,\dot{\varphi})$ plane (left) and in the $(\psi,\dot{\psi})$ plane (right) in the case $w_0=0$ and $V_0=0.03$. The analytical orbit $\varphi_1$ of Eq. \eqref{19} (black) is a repeller.}
\label{retratgrw0}
\end{figure}

In this scenario, 
 a fundamental quantity is the effective EoS parameter, namely
 \begin{eqnarray}\label{23} w=\frac{P}{\rho}=-1-\frac{2\dot{H}}{3H^2}=-1-\frac{\dot\rho}{3H\rho},
 \end{eqnarray}
which could be used to indicate the evolution of the EoS for each orbit of the dynamical system. In Figure \ref{weffgrw0}, we can see that all orbits tend asymptotically at the beginning of the contracting phase to $w=w_0$.

\begin{figure}[H]
\begin{center}
\includegraphics[scale=0.30]{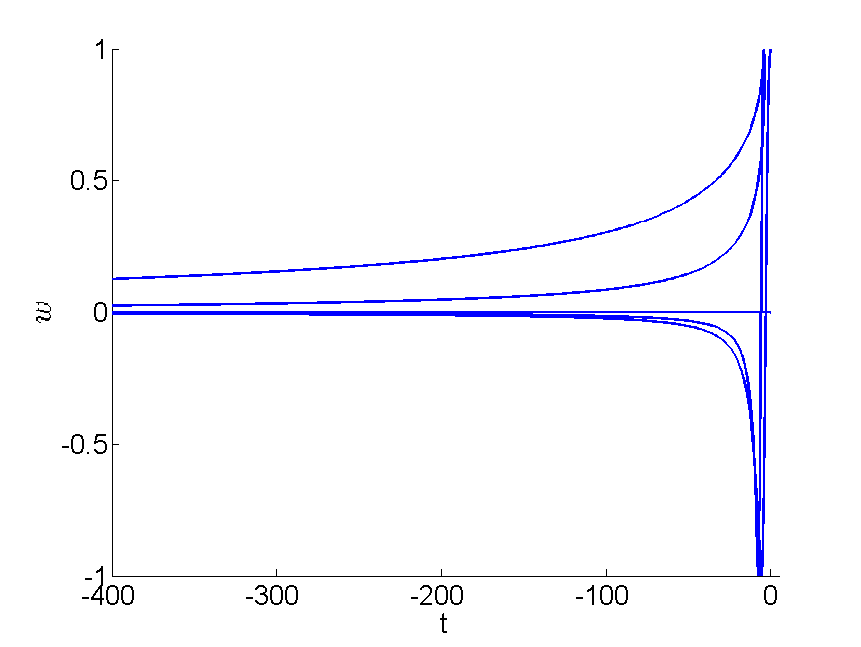}
\end{center}
\caption{Effective EoS parameter $w$ for orbits represented in the phase portrait in Figure \ref{retratgrw0}, { i.e., for $w_0=0$. One can see that going backward in time, for all orbits, $w$ converges to
zero}.}
\label{weffgrw0}
\end{figure}

\item $w_0>1$.
The system is only defined for
$|\dot\psi|\geq \sqrt{\frac{3(1+w_0)|V_0|}{2}}$. Moreover,
the system has three fixed points, namely
\begin{eqnarray}\label{22}\dot\psi_{1}=-(1+w_0)\sqrt{\frac{3|V_0|}{2|1-w_0|}}, \quad \mbox{and}\quad \dot\psi_{2,\pm}=\pm\sqrt{\frac{3(1+w_0)|V_0|}{2}}.
\end{eqnarray}

$\dot\psi_{1}$ is an attractor in the half plane $\dot\psi<0$ because $F(-\infty)=+\infty$ and $\dot\psi_{2,\pm}$ are repellers because these solutions 
correspond to $H=0$, and in the contracting phase $H$ decreases as time 
goes forward, that is, the orbits moves away from $\dot\psi_{2,\pm}$.

The first solution corresponds to
\begin{eqnarray}\varphi_{1}(t)=- \frac{1}{\sqrt{3(1+w_0)}}\ln\left({\frac{3V_0(1+w_0)^2}{2(1-w_0)}}t^2\right),
\end{eqnarray}
whose orbit depicts, all the time, in the contracting phase, a Universe with EoS $P=w_0\rho$.  The other ones satisfy $H=0$.

Since the orbit $\gamma_1(t)=(\varphi_{1}(t), \dot\varphi_{1}(t))$ is an attractor in the half plane $\dot\varphi>0$, this means that all the other orbits in the half plane $\dot\varphi>0$ depict, at late time, a Universe with EoS $P=w_0\rho$.

\begin{figure}[H]
\begin{center}
\includegraphics[scale=0.34]{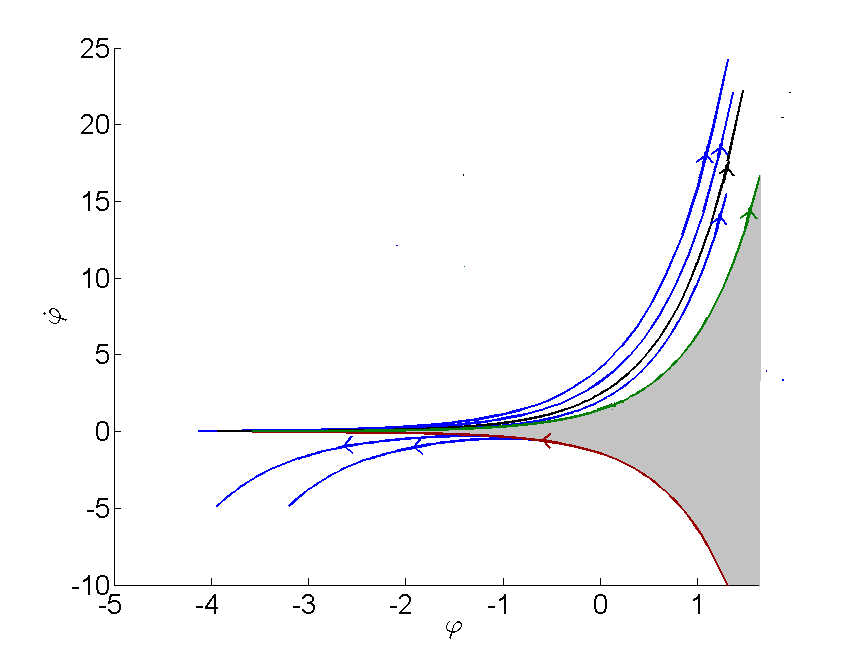}
\includegraphics[scale=0.27]{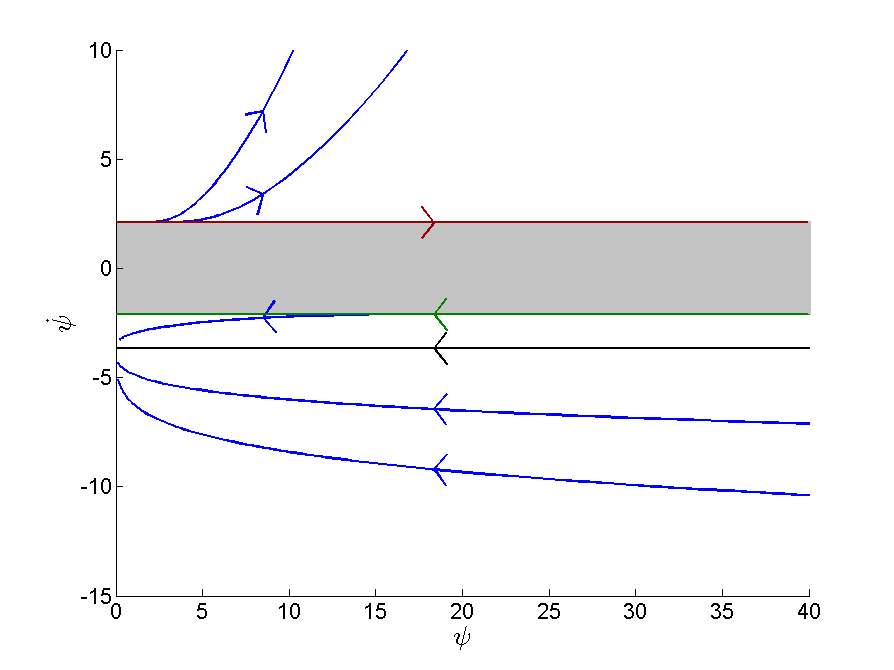}\end{center}

\caption{Phase portrait in the $(\varphi,\dot{\varphi})$ plane (left) and in the $(\psi,\dot{\psi})$ plane (right) in the case $w_0=2$ and $V_0=-1$. The analytical orbits
of Eq. \eqref{22} define the orbits $\varphi_1$ (black), $\varphi_{2,+}$ (red),
$\varphi_{2,-}$ (green). Nearby orbits (blue) show how $\varphi_1$ is an attractor
for the orbits between it and $\varphi_{2,-}$ (corresponding to the region 
where $\dot{\varphi}>0$ in phase space), while $\varphi_{2,+}, \varphi_{2,-}$ are 
repellers.}
\label{retratgrwgt1}
\end{figure}


Note that in Figure \ref{retratgrwgt1}, orbits approaching either of the fixed points $\dot{\psi}_{2,\pm}$ reach such value (corresponding to $\rho=0$) in a finite cosmic time. Since near $\dot{\psi}_{2,\pm}$ the dynamical system behaves as $\frac{d\dot{\psi}}{d\varphi}=-\frac{3}{2}\sqrt{1+w_0}\sqrt{\frac{2\dot{\psi}^2}{3(1+w_0)}+V_0}$, it is reached in a finite $\varphi$ time. Given that $\varphi$ is finite, $\dot{\varphi}\left(=\sqrt{-2V(\varphi)}\right)$ is finite as well and, hence, in a finite cosmic backward time the contracting phase starts after $\rho=0$. { This means that these orbits can be extended backward in time to the expanding phase. Hence, going forward in time one can argue that these orbits depict a bouncing universe, but this is not a good bounce, because the universe moves from the expanding to the contracting phase and, moreover, the
bounce occurs at $\rho=0$. As we will see in next Section dealing with Loop Quantum Cosmology, the bounces we are interested in are those that come from the contracting to expanding phase and such that at the bouncing time the energy density is very high. }
{ 
On the other hand, the orbits such that $\dot{\psi}_0\leq \dot{\psi}_1$ 
start from an initial value corresponding to $H=0$, which takes places at $t\to-\infty$  and evolve towards $H\to-\infty$.}

\

In Figure \ref{weffgrwgt1} we show the evolution of $w$. We see that for those orbits approaching to the analytical solution in the plane $(\psi,\dot{\psi})$, { at late times} $w$ tends to $w_0$, while for those in which $\dot{\psi}$ diverge, $w$ tends to $1$. { Regarding early times, those orbits coming from an infinite value of $w$ are those that have started the contracting phase for a finite value of time in the values $\dot{\psi}_{2,\pm}$, while the orbits that have started the contracting phase for $t\to-\infty$ approach backwards in time to the value $w=1$. }

\begin{figure}[H]
\begin{center}
\includegraphics[scale=0.40]{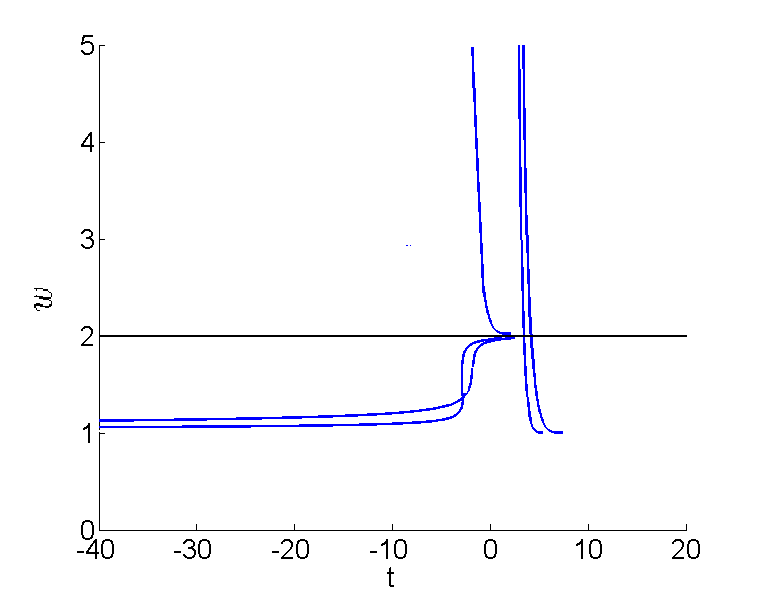}
\end{center}
\caption{Effective EoS parameter $w$ for the orbits represented in the phase portrait in Figure \ref{retratgrwgt1}.}
\label{weffgrwgt1}
\end{figure}

\end{enumerate}

\vspace{0.5cm}
\begin{remark}
The same analysis shows that, in the expanding phase, for $w_0=1$ there are solutions that correspond to a stiff fluid, and others with $H=0$. For $-1<w_0<1$ the solution 
that depicts a fluid with EoS $P=w_0\rho$ is an attractor. For $w_0>1$, the solution that depicts a fluid
with EoS $P=w_0\rho$ is a repeller, and the one with $H=0$ an attractor.
\end{remark}

 Once we have performed this previous study, it is interesting to consider the matter-ekpyrotic scenario given, in the contracting (resp. expanding) phase, by a potential ($w_0>1$)
 with the shape 
 \begin{eqnarray}\label{superposGR}
  V(\varphi)=V_0 e^{\sqrt{3}\varphi}+V_1e^{\sqrt{3(1+w_0)}\varphi},  \left(\mbox{resp.} \quad V(\varphi)=V_0 e^{-\sqrt{3}\varphi}+V_1e^{-\sqrt{3(1+w_0)}\varphi}\right)
 \end{eqnarray}
with $V_0>0$ and $V_1<0$.

\

\begin{figure}[H]
\begin{center}
\includegraphics[scale=0.27]{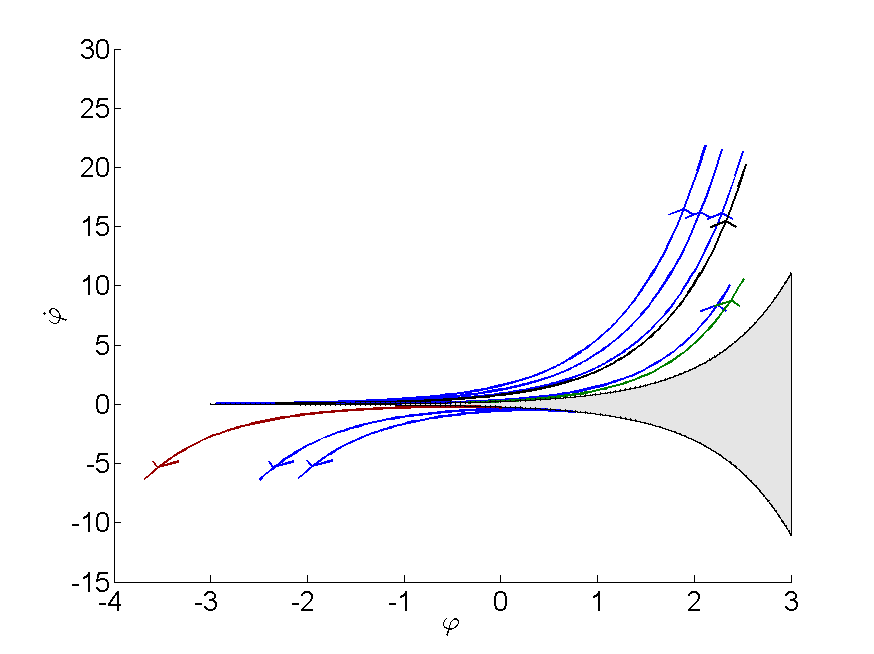}
\includegraphics[scale=0.27]{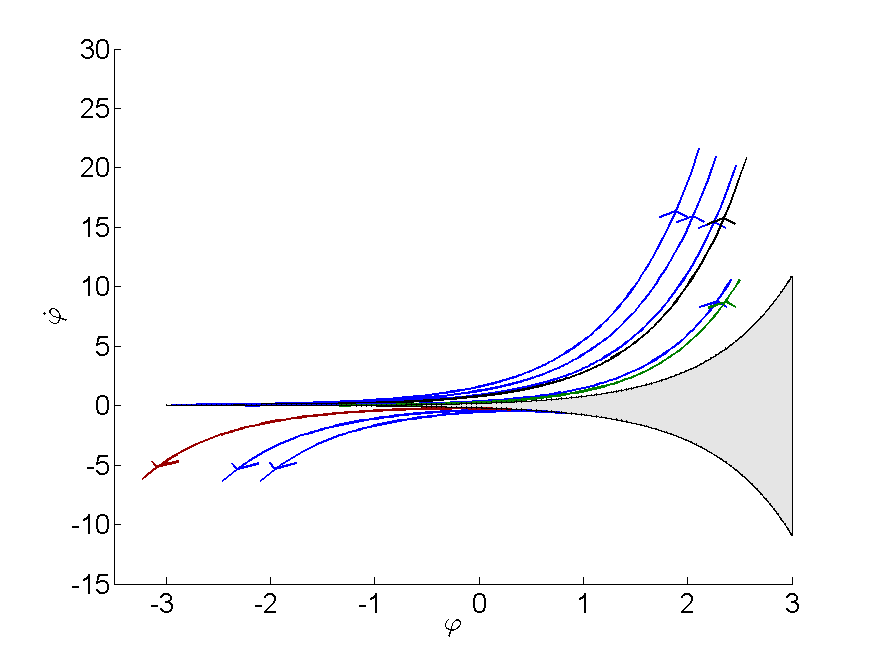}
\includegraphics[scale=0.28]{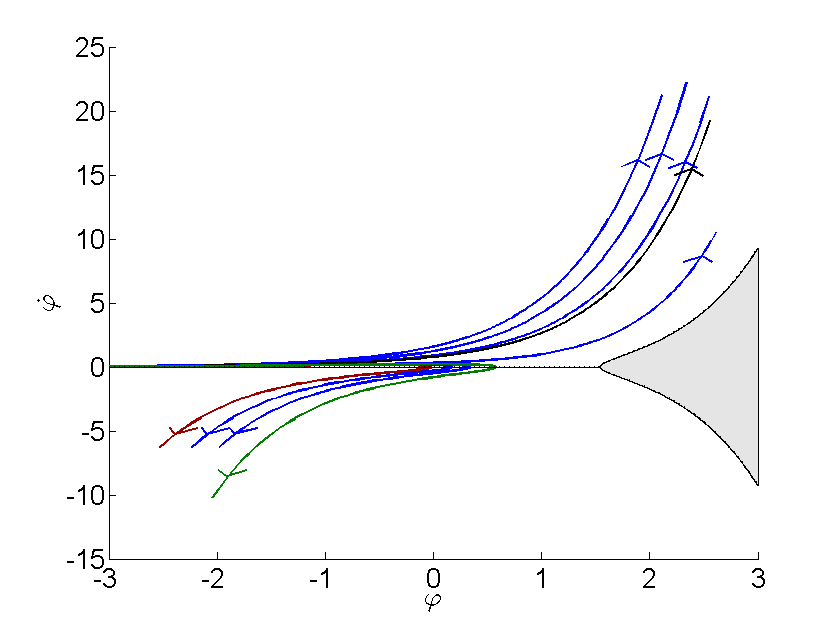}
\includegraphics[scale=0.27]{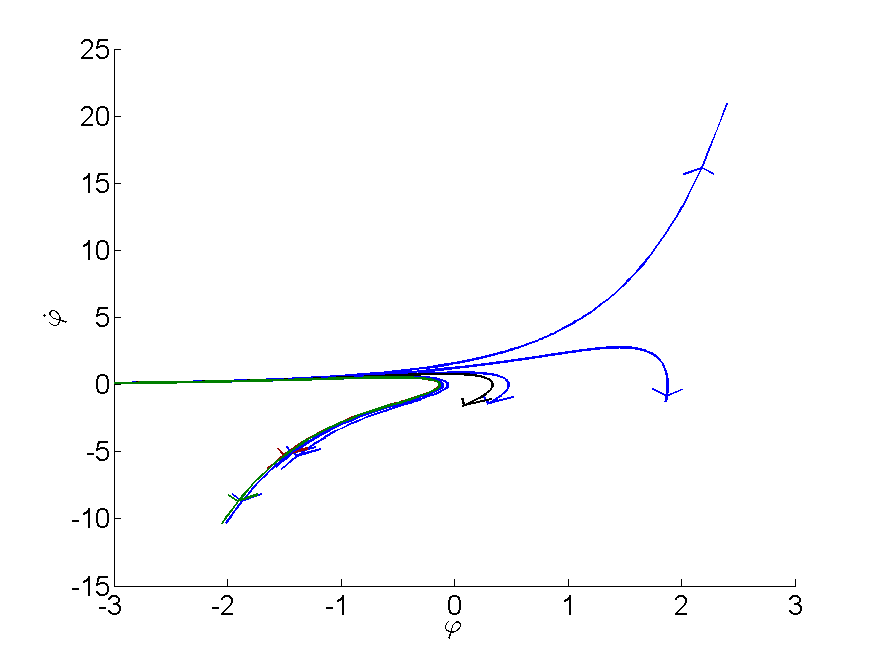}
\end{center}
\caption{Distinguished orbits in the $(\varphi,\dot{\varphi})$ plane for the superposed 
potential \eqref{superposGR} in the case $w_0=1.2$ and $V_1=-0.03$, with $V_0=10^{-3},10^{-2},10^{-1},1$
respectively.}
\label{fig:superposGR}
\end{figure}

\begin{figure}[H]
\begin{center}
\includegraphics[scale=0.30]{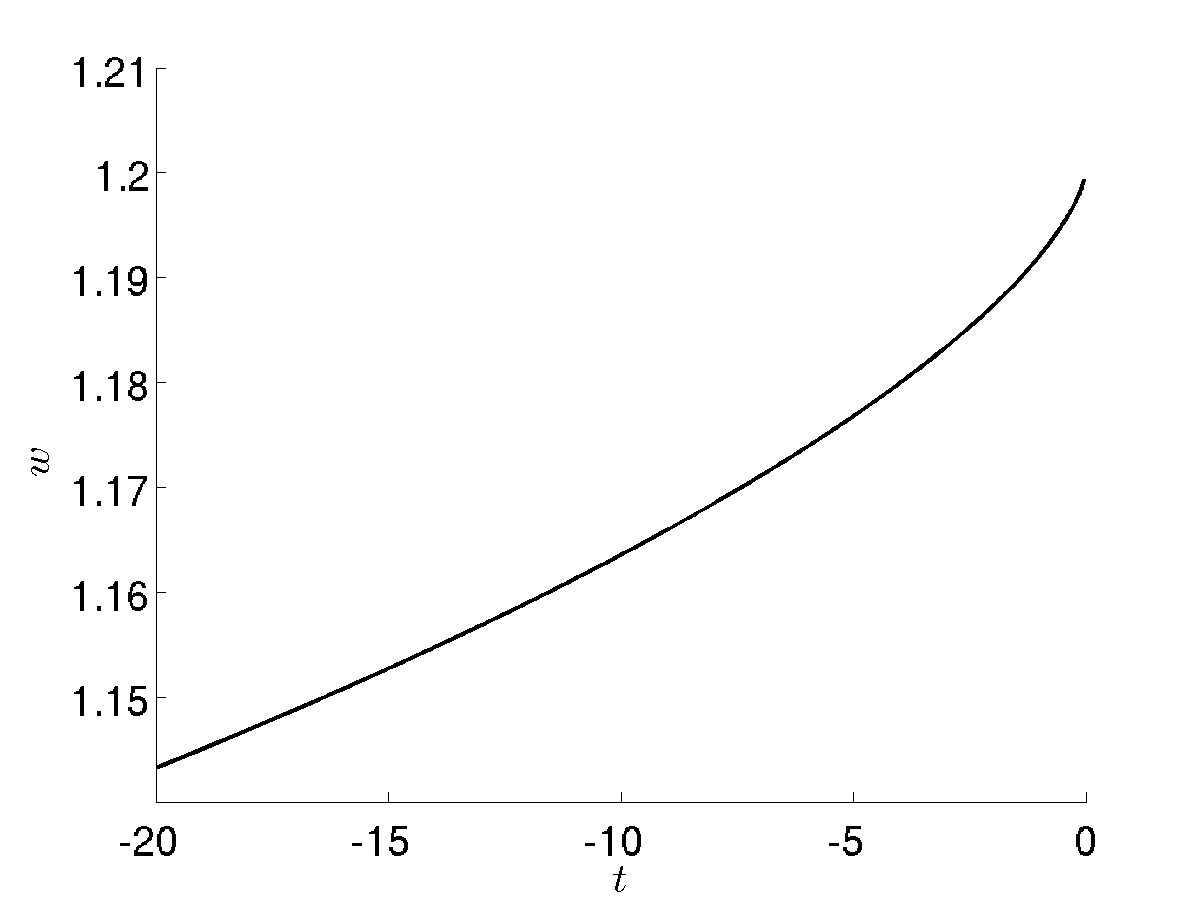}
\includegraphics[scale=0.30]{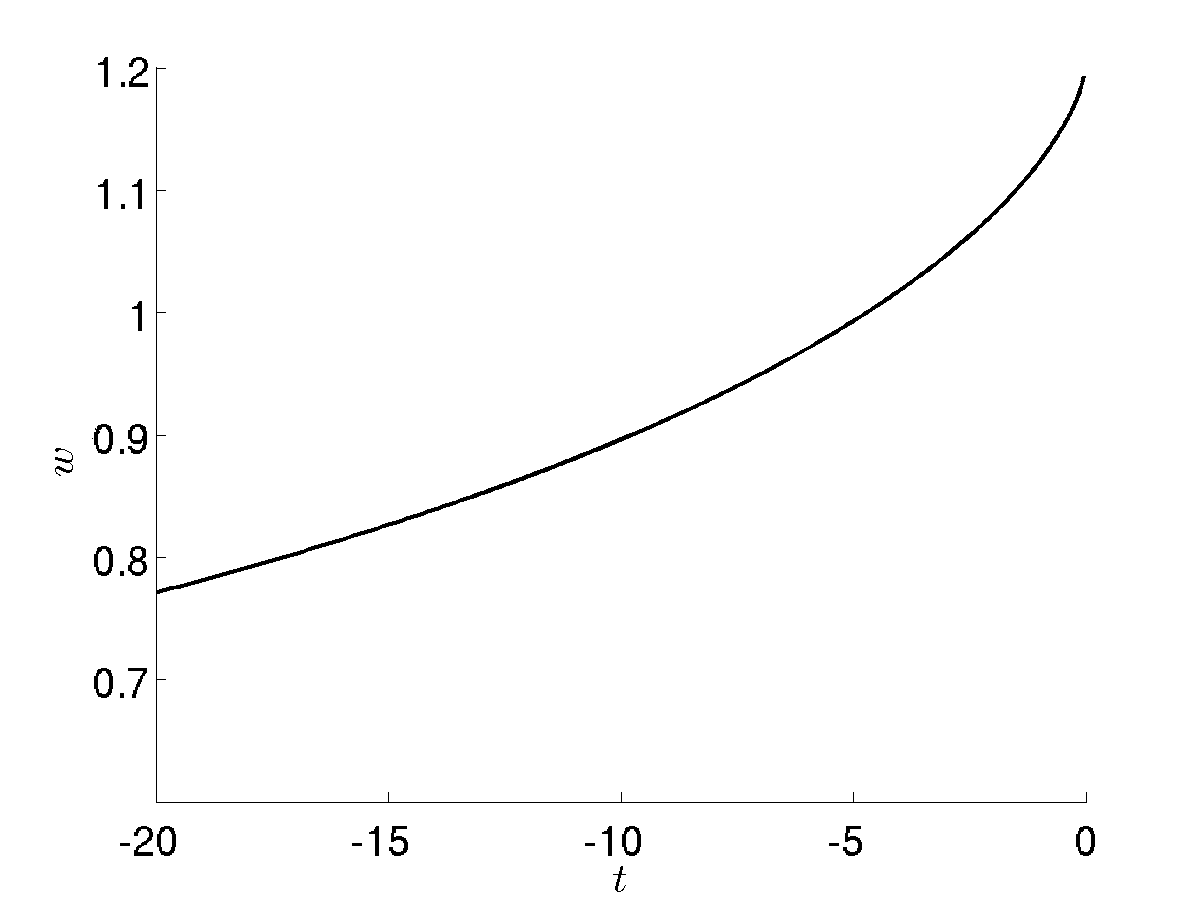}
\includegraphics[scale=0.30]{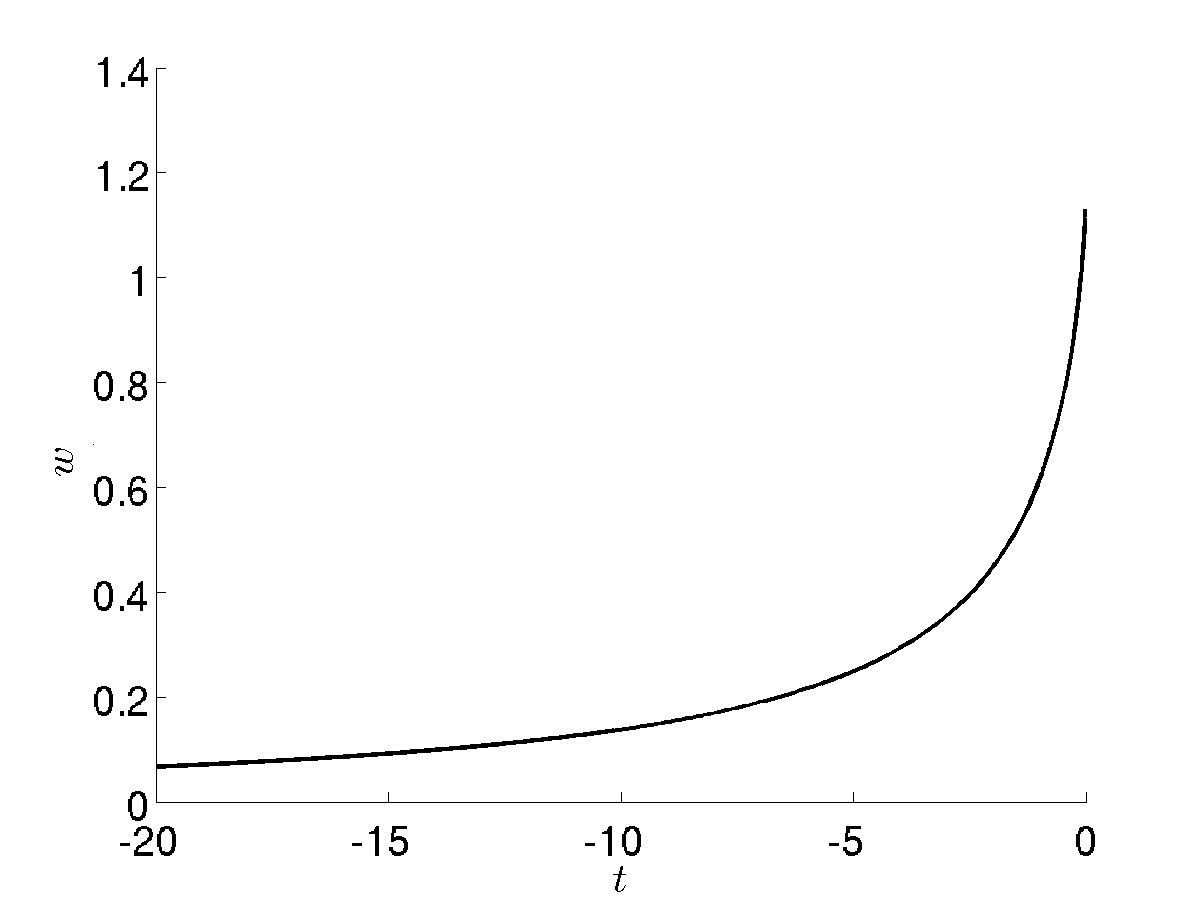}
\includegraphics[scale=0.27]{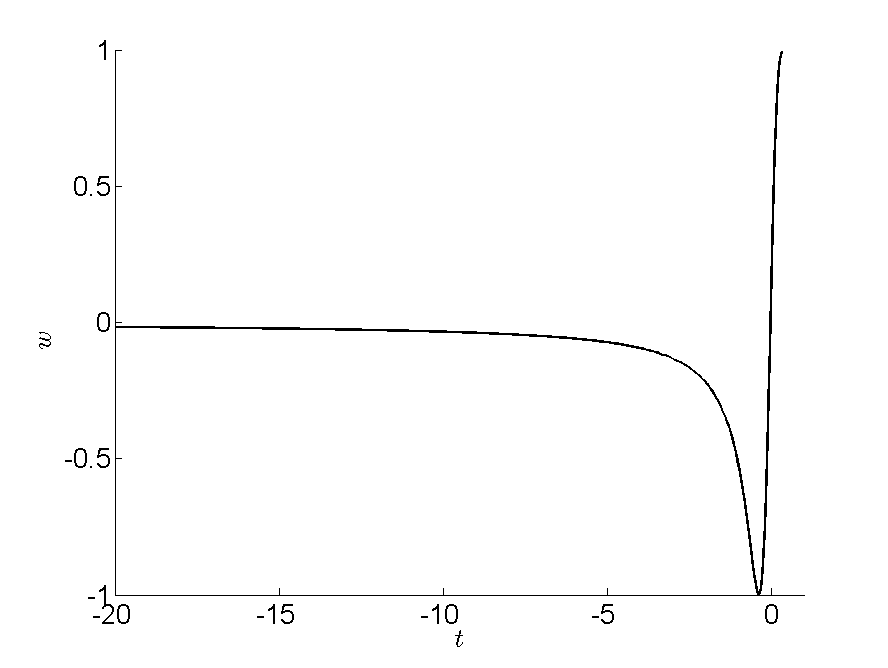}
\end{center}
\caption{The effective EoS parameter $w$ for the distinguished orbits 
indicated in black in Figure \ref{fig:superposGR}, again with $V_0=10^{-3},10^{-2},10^{-1},1$
respectively. One can see that for all the orbits $w$ starts at $0$ and ends at $w_0$.}
\label{fig:weffsuperposGR}
\end{figure}

In Figure \ref{fig:superposGR} we show phase portraits, in contracting phase, 
for the superposed potential \eqref{superposGR}
with $V_1=-0.03$ and $V_0=10^{-3},10^{-2},10^{-1},1$. The colouring code is 
that the black, red, green solutions are those given by the initial 
condition of the corresponding critical orbits in the case $V_0=0$. Moreover, in
Figure \ref{fig:weffsuperposGR}  we have drawn the parameter $w$ for 
black 
orbits in each superposition with $V_0=10^{-3},10^{-2},10^{-1},1$.

\

 From our previous study it seems clear that in the contracting phase, at early times, all the orbits will depict a matter dominated Universe, and after leaving this behaviour 
 they will evolve, at late times, to an ekpyrotic one. 
 On the contrary, in the expanding phase, at early times, orbits will depict a Universe in the ekpyrotic phase that eventually   evolves to a matter dominated one.

 \section{The matter-ekpyrotic scenario in Loop Quantum Cosmology: The background}

 {  
 Basically, Loop Quantum Cosmology (LQC), where holonomy corrections are introduced due to  the  assumption of the discrete nature of the space-time,  could be built following three steps: 
 \begin{enumerate}
 \item In GR, for the flat FLRW geometry,  one takes a cell of volume $1$, the gravitational part of the Hamiltonian is  ${\mathcal H}=-3H^2a^3$, and the usual canonical conjugate variables are  the pair $(a, p_a=-3\dot{a}a)$. In LQC
 these canonically conjugate variables  are  replaced by the Ashtekar connection and its conjugate momentum (the densitized triad), that we will denote respectively by $c$ and $p$, and which are related with
 the standard variables via the relation \cite{cm}
 \begin{eqnarray}
 c=\gamma \dot{a}\qquad p=a^2,
 \end{eqnarray}
 where $\gamma\cong 0.2375$ is the Barbero-Immirzi parameter, whose numerical value is obtained comparing the Bekenstein-Hawking formula with the
 black hole entropy calculated in Loop Quantum Gravity (LQG) \cite{miessner}. The Poisson bracket for these variables satisfies $\{c,p\}=\frac{\gamma}{3}$. Moreover, in LQC another pair of canonically conjugate
 variables  $(\beta, {\mathcal V})$,  named new variables,  satisfying 
 $\{\beta,{\mathcal V}\}=\frac{\gamma}{2}$ 
and defined as
 $\beta\equiv \frac{c}{p^{1/2}}=\gamma H$ and ${\mathcal V}\equiv p^{3/2}=a^3$, are usually used.

 \item The quantization of GR via the Wheeler-DeWitt equation  in the $a$-representation is based on the Hilbert space ${\mathcal L}^2(\R^+, da)$, which shows the continuous nature of the space (the scale factor has a continuous spectrum). However, in LQC the Hilbert space is the space of almost periodic functions, i.e.,
 functions of the form \cite{abl}
 \begin{eqnarray}\Psi(c)=\sum_{n\in\N} \alpha_n e^{\frac{i\mu_n c}{2}}
 \end{eqnarray}  with 
 $\mu_n\in \R$ and $\alpha_n$ being a square-summable sequence ($\sum_{n\in\N} |\alpha_n|^2<\infty$), provided by the inner product (in the $c$-representation)
 \begin{eqnarray}
 \langle \Phi, \Psi\rangle=\lim_{L\rightarrow \infty} \frac{1}{2L}\int_{-L}^L \Phi^*(c) \Psi(c) dc.
 \end{eqnarray} 
 
 The orthonormal basis of this space are functions of the form $|\mu \rangle \equiv  e^{\frac{i\mu c}{2}} $, and the quantum operator corresponding to the momentum $p$, in this representation,
 is given by $\hat{p}=-\frac{i\gamma }{3}\frac{d}{dc}$. Thus, the ortonormal states $|\mu\rangle$ are eigenvectors of $\hat{p}$, which proves that its spectrum is discrete, showing the discrete nature of the space in LQC \cite{bojowald}. However, the variable $c$ (and also $\beta$) does not have a well-defined quantum operator $\hat{c}$, because the norm of $\hat{c}|\mu\rangle$ diverges. For this reason it is important to introduce holonomies  \cite{ashtekar} obtained by integrating the Ashtekar connection along a curve 
  $h_j(\mu)=e^{-i\frac{\mu c}{2}\sigma_j}
  =\cos(\frac{\mu c}{2} )-i\sigma_j\cos(\frac{\mu c}{2} )$, where $\mu$ depends on the length of the curve and which has a well-defined quantum operator in the Hilbert space.
 
 \item In these variables the gravitational part of the Hamiltonian becomes
 \begin{eqnarray}
 {\mathcal H}=-\frac{3}{\gamma^2}c^2p^{1/2}\qquad \mbox{or} \qquad {\mathcal H}=-\frac{3}{\gamma^2}\beta^2{\mathcal V},
 \end{eqnarray}
 which contains the variable $c$ or $\beta$, and thus it does not have a well-defined quantum analogue. Then, and this is a key point in LQC, one has to ``regulate" (in fact, this means approximate) the Hamiltonian, obtaining another one that contains quantities that could be quantized. Working in the $\beta$-representation this could be done using the holonomies
 $h_j(\lambda)=e^{-i\frac{\lambda \beta}{2}\sigma_j}
  =\cos(\frac{\lambda \beta}{2} )-i\sigma_j\cos(\frac{\lambda \beta}{2} )$,  where now $\lambda$ is a parameter with the dimensions of length, whose numerical value is obtained, invoking the quantum nature of the space, 
  identifying its square with the minimum eigenvalue of the area operator in Loop Quantum Gravity, namely $\lambda=\sqrt{\frac{\sqrt{3}}{2}\gamma}$ \cite{Psingh}.
  This regularized gravitational part of the Hamiltonian has a complicated expression in terms of the holonomies \cite{ashtekar, bojowald, abl}, but after some algebra it reduces to \cite{he1, dmp}
  \begin{eqnarray}
  {\mathcal H}_{LQC}=-3{\mathcal V}\frac{\sin^2(\lambda\beta)}{\lambda^2\gamma^2}\Longrightarrow 
  {\mathcal H}_{T}=-3{\mathcal V}\frac{\sin^2(\lambda\beta)}{\lambda^2\gamma^2} +\rho {\mathcal V}, \end{eqnarray}
  where ${\mathcal H}_T$ denotes the total Hamiltonian of the system. Note also that, when $\lambda$ goes to zero, it becomes the usual Hamiltonian in GR.
  
  Finally,  the Hamilton equation $\dot{\mathcal V}=\{V,H_T\}$ is equivalent to the equation
  \begin {eqnarray}
  H^2=\frac{\sin^2(\lambda\beta)(1-\sin^2(\lambda\beta))}{\lambda^2\gamma^2},
  \end{eqnarray}
 that together with the Hamiltonian constraint, i.e., $ \frac{\sin^2(\lambda\beta)}{\lambda^2\gamma^2} =\rho/3$, leads to the Friedmann equation in LQC
 \cite{singh}
 \begin{eqnarray} \label{25} H^2=\frac{\rho}{3}\left(1-\frac{\rho}{\rho_c} \right),
 \end{eqnarray}
where $\rho_c\equiv\frac{3}{\lambda^2\gamma^2}=\frac{2\sqrt{3}}{\gamma^3}\cong 258$ is the so-called critical energy density.

It is important to realize that this equation depicts an ellipse in the plane $(H,\rho)$, which is a bounded curve, meaning that the Big Bang singularity is removed in this theory. In fact,
it is replaced by a Big Bounce when the universe reaches the critical energy density $\rho_c$. Moreover note that the energy density lies between $0$ and $\rho_c$ and the Hubble parameter
between $-\rho_c/12$ and $\rho_c/12$. This is a great difference in contrast with GR, where the Friedmann equation depicts a parabola in the plane $(H,\rho)$ allowing the Big Bang singularity because this curve is unbounded.

\end{enumerate}
}

 Once again, we want to mimic by a scalar field  a fluid with EoS $P=w_0\rho$. Solving the Friedmann and conservation equations in LQC for this kind of fluid
 one has the solution \cite{wilson}

 \begin{eqnarray} H(t)=\frac{1+w_0}{2}t\rho(t),\quad \rho(t)=\frac{\rho_c}{\frac{3}{4}(1+w_0)^2\rho_ct^2+1}.
 \end{eqnarray}

 Then, the equation $P=w_0\rho$ becomes
 \begin{eqnarray}\dot{\varphi}^2=(1+w_0)\rho=\frac{(1+w_0)\rho_c}{\frac{3}{4}(1+w_0)^2\rho_ct^2+1},
 \end{eqnarray}
 whose solution is given by \cite{Haro3}
 \begin{eqnarray}\label{28}\varphi(t)=\frac{2}{\sqrt{3(1+w_0)}}\ln\left(\frac{\sqrt{\frac{3}{4}(1+w_0)^2\rho_c}t+\sqrt{\frac{3}{4}(1+w_0)^2\rho_ct^2+1}}{|\varphi_0|}  \right).
 \end{eqnarray}

 To reconstruct the potential we use the formula
 \begin{eqnarray}V=\frac{1-w_0}{2}\rho=\frac{1-w_0}{2}\frac{\rho_c}{\frac{3}{4}(1+w_0)^2\rho_ct^2+1},
 \end{eqnarray}
 and we isolate $\frac{3}{4}(1+w_0)^2\rho_ct^2+1$ as a function of $\varphi$, obtaining \cite{wilson}
 \begin{eqnarray}\label{eq:Vlqc}
 V(\varphi)=V_1\frac{e^{\sqrt{3(1+w_0)}\varphi}}{\left(1+\frac{V_1}{2\rho_c(1-w_0)}e^{\sqrt{3(1+w_0)}\varphi}\right)^2},
 \end{eqnarray}
 where we have chosen $\varphi_0^2=\frac{V_1}{2\rho_c(1-w_0)}$.

\
 
We can see, as in the case of General Relativity,  for $w_0=1$ the potential vanishes, it is positive for $-1<w_0<1$ and negative for $w_0>1$. As in General Relativity, we will differentiate between $3$ cases so as to study the following dynamical system:
\begin{eqnarray}
\rho=\frac{\dot{\varphi}^2}{2}+V(\varphi)\quad \ddot{\varphi}+3H_{\pm}\dot{\varphi}+V_{\varphi}=0,\end{eqnarray}
where $H_{\pm}=\pm\sqrt{\frac{\rho}{3}\left(1-\frac{\rho}{\rho_c} \right)}$.

\begin{enumerate}

\item $w_0=1$. In this case the differential equation becomes:
{
\begin{eqnarray}
\ddot{\varphi}=\mp \sqrt{\frac{3}{2}}|\dot{\varphi}|\dot{\varphi}\sqrt{1-\frac{\dot{\varphi}^2}{2\rho_c}}
\end{eqnarray}}
obtaining for $\dot{\varphi}>0$ orbits in the plane $(\varphi,\dot{\varphi})$ of the form
\begin{eqnarray}
\gamma_+(t)=\left(\varphi_0+\sqrt{\frac{2}{3}}\ln\left(\sqrt{3\rho_c}t+C+\sqrt{1+(\sqrt{3\rho_c}t+C)^2}\right), \sqrt{\frac{2\rho_c}{1+(\sqrt{3\rho_c}t+C)^2}} \right),
\end{eqnarray}
where $C=\sqrt{\frac{2\rho_c}{\dot{\varphi_0}^2}-1}$, and for $\dot{\varphi}<0$
\begin{eqnarray}
\gamma_-(t)=\left(\varphi_0-\sqrt{\frac{2}{3}}\ln\left(\sqrt{3\rho_c}t+D+\sqrt{1+(\sqrt{3\rho_c}t+D)^2}\right), -\sqrt{\frac{2\rho_c}{1+(\sqrt{3\rho_c}t+D)^2}} \right),
\end{eqnarray}
where $D=-\sqrt{\frac{2\rho_c}{\dot{\varphi_0}^2}-1}$. { And finally the trivial solution from the previous equation would correspond to $\gamma_0(t)= (\sqrt{2\rho_c}t+C,\sqrt{2\rho_c})$, corresponding to $H=0$.

Now, we proceed to the stability analysis analogous to the one performed in the case of GR. If we consider for instant in 
{$\gamma_+(t)$}  a small perturbation $\tilde{C}=C+\delta C$, then one obtains that 
{
$$\frac{\delta X(t)}{X(t)}=\frac{(C+\sqrt{3\rho_c}t)\delta C}{1+(\sqrt{3\rho_c}t+C)^2}\leq \frac{\delta C}{\sqrt{1+(\sqrt{3\rho_c}t+C)^2}}\leq \delta C.$$
Moreover, we observe that 
$\frac{\delta X(t)}{X(t)}$} tends to 0 both at { $t=-C/\sqrt{3\rho_c}$ }
(i.e the bounce, the end of contracting phase) and at $t\to\infty$ (end of expanding phase). Therefore, all orbits are stable and, furthermore, all orbits foliating the phase space satisfy throughout all the evolution of time the equation of state $P=\rho$.
}

\

In order to study the remaining two cases, it is suitable to perform the following change of variables, motivated by \eqref{28}:
\begin{eqnarray}
\psi=\sinh\left(\frac{\varphi\sqrt{3(1+w_0)}}{2}+\frac{1}{2}\ln\left(\frac{V_1}{2\rho_c(1-w_0)}\right)\right)
\end{eqnarray}

Hence, equation of conservation becomes
\begin{eqnarray}
\ddot{\psi}=-3H_{\pm}(\psi,\dot{\psi})\dot{\psi}+\rho(\psi,\dot{\psi})\psi\frac{3(1+w_0)}{2},
\end{eqnarray}
where $H_{\pm}=\pm\sqrt{\frac{\rho}{3}\left(1-\frac{\rho}{\rho_c} \right)}$ and $\rho=\frac{2}{3(1+w_0)(1+\psi^2)}\left(\dot{\psi}^2+\frac{3(1-w_0^2)}{4}\rho_c \right)$.

\item $-1<w_0<1$. The dynamical system is defined for $\dot{\psi}^2\leq \frac{3\rho_c(1+w_0)}{4}(1+w_0+2\psi^2)$ and has two fixed points:
\begin{eqnarray}
\dot{\psi}_1^{\pm}=\pm \frac{\sqrt{3\rho_c}}{2}(1+w_0)
\end{eqnarray}

In Figure \ref{retratlqcw0}, we have represented the phase portrait of the dynamical system, such that blue orbits stand for the analytical ones (fixed points $\dot{\psi}_1^{\pm}$, corresponding to equation \eqref{28}), red colour is for the contracting phase and green colour is for the expanding phase. The coloured zone in the phase portrait is the forbidden region, in which $\rho>\rho_c$. Note that the bounce from the contracting to the expanding phase happens in $\rho=\rho_c$ curve. We see that there are two types of orbits: those that cross axis $\varphi=0$ and those that cross axis $\dot{\varphi}=0$. Therefore, it can be clearly concluded that the analytical orbit is a repeller in the contracting phase and an attractor in the contracting phase, as one can also verify with the effective EoS parameter in Figure \ref{wefflqcw0}.
  \begin{figure}[H]
\begin{center}
\includegraphics[scale=0.34]{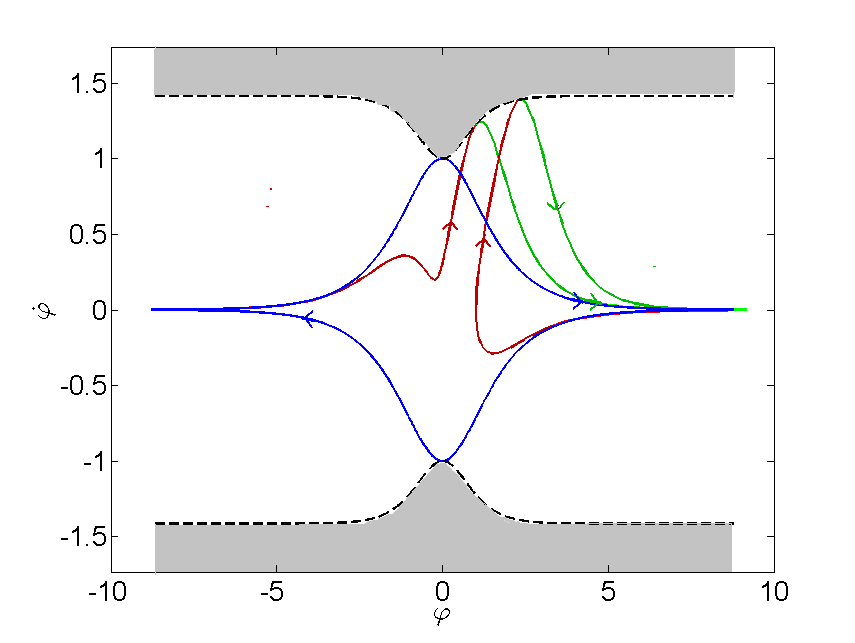}
\includegraphics[scale=0.35]{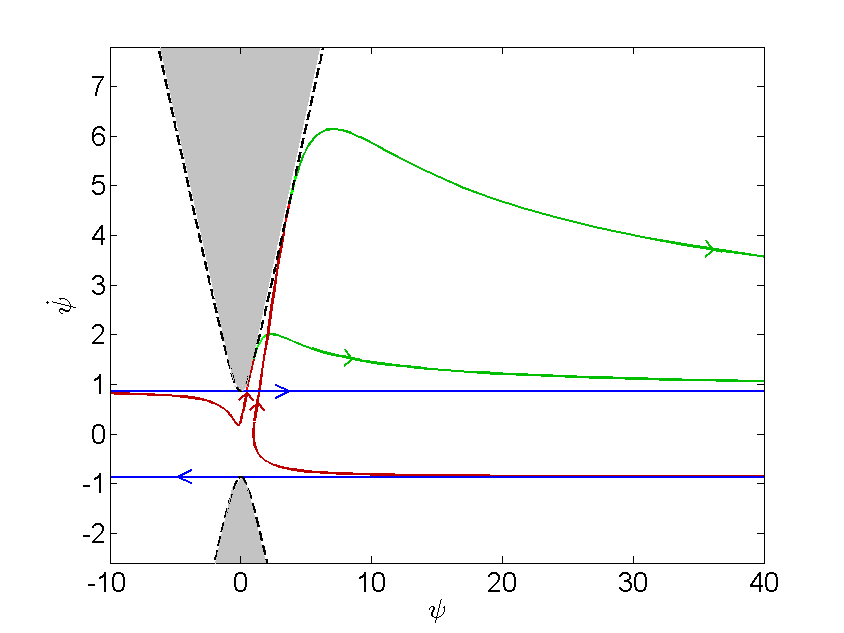}
\end{center}

\caption{ Phase portrait in the $(\varphi,\dot{\varphi})$ plane (left) and in the $(\psi,\dot{\psi})$ plane for the LQC system
with potential \eqref{eq:Vlqc} for $w_0=0$. The analytical solution \eqref{28} has been outlined in blue.}
\label{retratlqcw0}
\end{figure}

\begin{figure}[H]
\begin{center}
\includegraphics[scale=0.30]{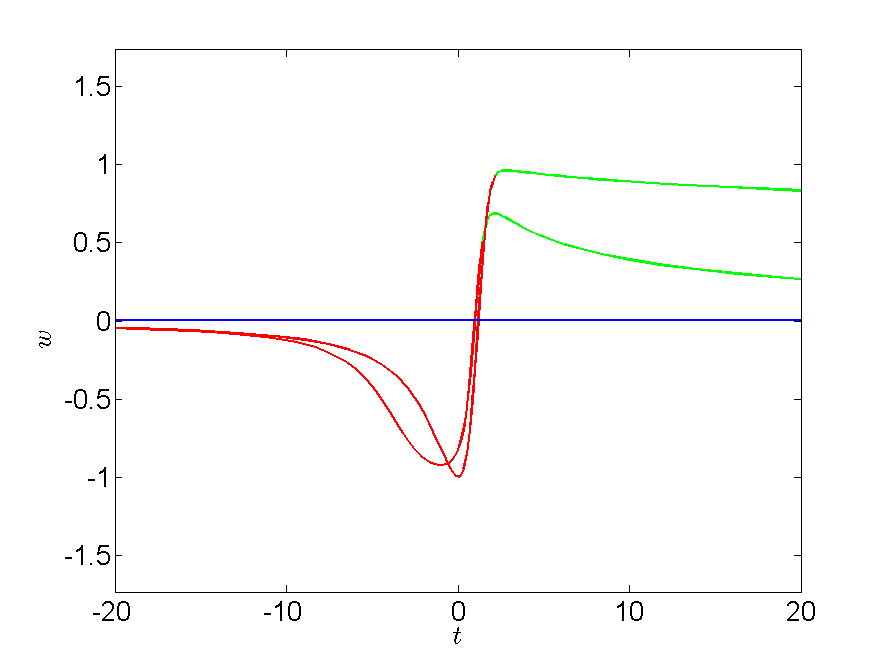}
\end{center}
\caption{Effective EoS parameter for the orbits shown in Figure \ref{retratlqcw0}.  The orbit that crosses $\dot{\psi}=0$ is the one which passes through $w=-1$. Both tend asymptotically at $t\to\infty$ and $t\to-\infty$ to $w_0$. }
\label{wefflqcw0}
\end{figure}

\item $w_0>1$. The dynamical system is defined for $\frac{3\rho_c(w_0^2-1)}{4}\rho_c\leq \dot{\psi}^2\leq \frac{3\rho_c(1+w_0)}{4}(1+w_0+2\psi^2)$ and has four fixed points:
\begin{eqnarray}
\dot{\psi}_1^{\pm}=\pm \frac{\sqrt{3\rho_c}}{2}(1+w_0) \quad \dot{\psi}_0^{\pm}=\pm \frac{\sqrt{3\rho_c}}{2}(1+w_0)
\end{eqnarray}

As already happened with General Relativity for the $w_0>1$ case, those orbits approaching $\dot{\psi}_0^{\pm}$ (corresponding to $\rho=0$) reach it at a finite time because of the same argument that was used in General Relativity, since holonomic corrections can be discarded near $\rho=0$. Therefore, in these cases we will see a bounce at $\rho=0$. In the phase portrait in Figure \ref{retratlqcwgt1} (with the same colour notation as in the former case), we have seen that all orbits different from the analytical one ($\dot{\psi}_1^{\pm}$, corresponding to \eqref{28}) perform two cycles in the $(H,\rho)$ ellipse, with two bounces at $\rho=\rho_c$ and one bounce at $\rho=0$. The process is the following:

\begin{itemize}

\item Contracting phase 1: At $t\to-\infty$, $\rho=0$ for infinite values of $\psi$ and $\dot{\psi}$ and at a finite time $t=t_1^b$ the orbit reaches $\rho=\rho_c$, occurring a bounce.

\item Expanding phase 1: After the first bounce, the orbit enters into the expanding phase reaching either $\dot{\psi}_0^+$ or $\dot{\psi}_0^-$ (corresponding to $\rho=0$) at a finite time $t=t_2^b$. There, another bounce occurs.

\item Contracting phase 2: After the second bounce, the orbit starts the second cycle in the ellipse $(H,\rho)$ and reaches $\rho=\rho_c$ at a finite time $t=t_3^b$ (third bounce).

\item Expanding phase 3: After this last bounce, the orbit enters into the expanding phase and tends asymptotically for $t\to\infty$ to $\rho=0$ for infinite values of both $\psi$ and $\dot{\psi}$.

\end{itemize}

Therefore, we infere that the analytical solutions $\dot{\psi}_1^{\pm}$ are attractors in the contracting phase and repellers in the expanding phase, while the solutions $\dot{\psi}_0^{\pm}$ are repellers in the contracting phase and attractors in the expanding phase. With regards to the behaviour of $w$, represented in \ref{wefflqcwgt1}, it is seen that both at $t\to-\infty$ and $t\to\infty$, $w=1$, whereas it diverges at the bounce at $\rho=0$.

\begin{figure}[H]
\begin{center}
\includegraphics[scale=0.34]{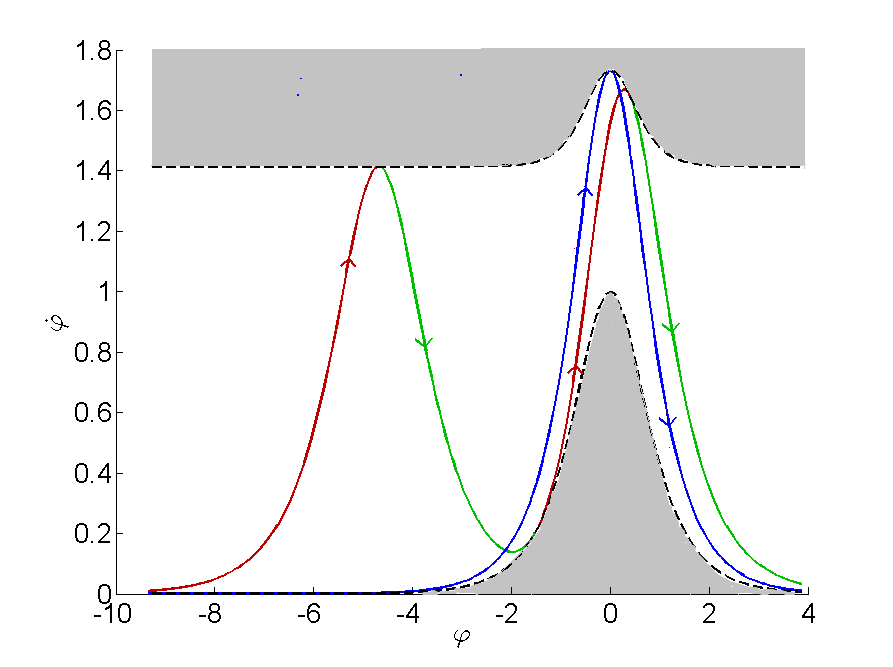}
\includegraphics[scale=0.34]{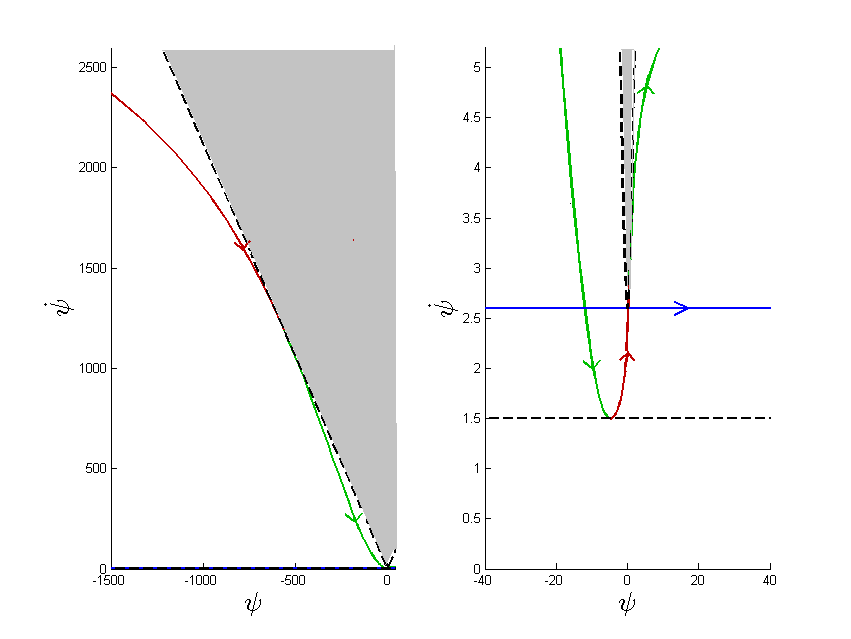}
\end{center}
\caption{ Phase portrait in the $(\varphi,\dot{\varphi})$ plane (left) and in the $(\psi,\dot{\psi})$ plane for the LQC system
with potential \eqref{eq:Vlqc} for $w_0=2$. The analytical solution \eqref{28} has been outlined in blue.}
\label{retratlqcwgt1}
\end{figure}

\begin{figure}[H]
\begin{center}
\includegraphics[scale=0.30]{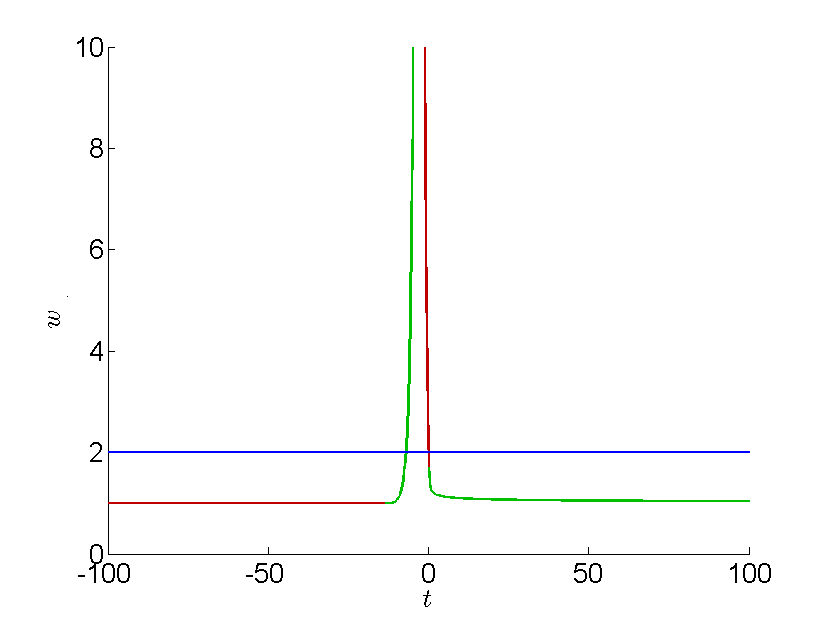}
\end{center}
\caption{Effective EoS parameter for the orbit shown in Figure \ref{retratlqcwgt1}. In the first cicle, $w$ starts being $1$ at very early times and diverges at the end of the cycle. In the second cycle, in the contracting phase, $w$ approaches to $w_0$ near the bounce and when the orbit enters in the expanding phase the effective EoS parameter goes asymptotically to $1$.}
\label{wefflqcwgt1}
\end{figure}

\end{enumerate}

\

From these previous results, to obtain a simple version of the matter-ekpyrotic bouncing scenario in LQC,  we can consider the potential 
 \begin{eqnarray}\label{POTENTIAL}
 V(\varphi)=\left\{\begin{array}{ccc}
 V_0\frac{e^{\sqrt{3}\varphi}}{\left(1+\frac{V_0}{2\rho_c}e^{\sqrt{3}\varphi}\right)^2} &\mbox{ for}& \varphi\leq \varphi_E\\
 & &\\
 V_1\frac{e^{\sqrt{3(1+w_0)}\varphi}}{\left(1+\frac{V_1}{2\rho_c(1-w_0)}e^{\sqrt{3(1+w_0)}\varphi}\right)^2}&\mbox{ for}& \varphi\geq \varphi_E, 
 \end{array}\right.
\end{eqnarray}
where $w_0>1$, $V_0>0$ and $V_1<0$ and the abrupt phase transition, needed in order to produce enough particles to reheat the universe, occurs at $\varphi=\varphi_E$.

For this potential some orbits will lead to an effective EoS being zero at early times, approximately $w_0$ just before the bounce, and finally approaching asymptotically to $1$ in the expanding phase. These orbits could be obtained by taking as an initial condition a point $(\varphi_0, \dot{\varphi}_0)$ corresponding to an orbit in the second cycle of the figure \ref{retratlqcwgt1}. Thus, with these initial conditions, integrating the conservation equation, one will set up, for the potential \eqref{POTENTIAL}, orbits with an effective EoS parameter satisfying the required properties. 

In fact, in last section we will obtain a potential which leads to a viable matter-ekpyrotic bounce scenario, i.e., a scenario which provides theoretical
 values of the spectral parameters that match at $2\sigma$ Confidence Level with the recent observational data.

\
 
Some final remarks are in order: 
 \begin{enumerate}\item
 As we have already explained in the introduction, holonomy corrections provide a Big Bounce. Then, the conservation equation 
 \begin{eqnarray}\label{30}
  \ddot{\varphi}+ 3H_{\pm}\dot{\varphi}+V_{\varphi}=0, 
 \end{eqnarray}
has to be studied in the contracting and expanding phase. Firstly, the contracting phase, i.e. with $H_{-}$, takes place and, when the Universe bounces at $H=0$, it enters into the expanding one and, thus, its dynamics will now be given by the conservation equation with $H_+$.

 \item
 Note also that, in Loop Quantum Cosmology, when one considers $w$ one has to use the last formula of \eqref {23}, i.e., 
 $w=-1-\frac{\dot\rho}{3H\rho}$, because the intermediate one only holds in General Relativity (when holonomy corrections can be disregarded).

 \item In \cite{wilson2}, the authors obtain an ekpyrotic phase with the following potential:
\begin{eqnarray}
V(\varphi)=-\frac{2V_0}{e^{-\sqrt{3(1+w_0)}\varphi}+e^{b_v\sqrt{3(1+w_0)}\varphi}} \label{potcwe}
\end{eqnarray}

When $b_v=1$, this potential is symmetric and, by using the change of variables $\psi=\sinh\left(\frac{\sqrt{3(1+w_0)}}{2}\varphi \right)$, the new potential becomes $V(\psi)=\frac{-V_0}{1+2\psi^2}$ which has the asymptotic behaviour as our former potential for $|\psi|\gg 1$.

If $V_0=\frac{\rho_c(w_0-1)}{2}$, the bounce at $\varphi=0$ ($\psi=0$) takes place at $w=w_0$. Therefore, in this case, all orbits have two cycles except the one bouncing at $w=w_0$, which has a single cycle, as we can see in Figure \ref{cairebot}. However, this last orbit does not correspond to the analytical solution since the value of $w$ does not remain constant, but approaches asymptotically to 1 for both $t\to\infty$ and $t\to-\infty$, as we can see in Figure \ref{cairebotweff}.

\begin{figure}[H]
\begin{center}
\includegraphics[scale=0.34]{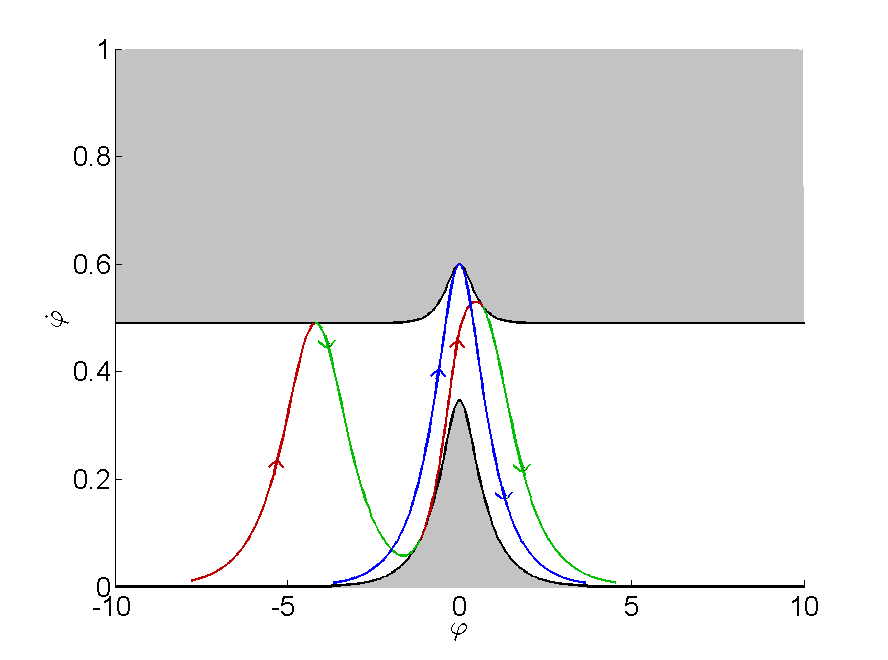}
\includegraphics[scale=0.34]{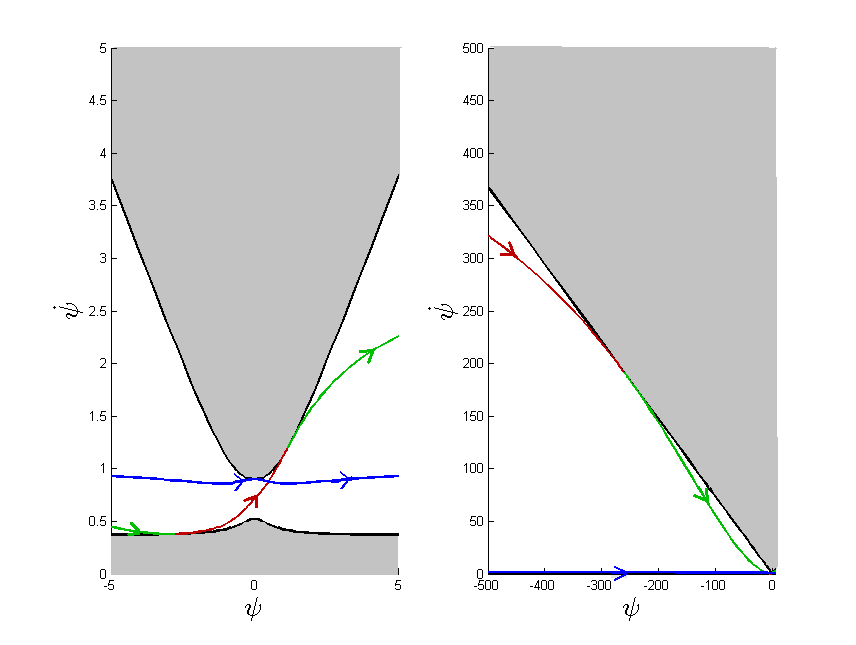}
\end{center}
\caption{Phase portrait in the $(\varphi,\dot{\varphi})$ plane (left) and in the $(\psi,\dot{\psi})$ plane for the LQC system
with potential \eqref{potcwe} for $w_0=2$ and $V_0=\frac{(w_0-1)\rho_c}{2}$. The orbit with the bounce at $w=w_0$ has been outlined in blue.}
\label{cairebot}
\end{figure}

\begin{figure}[H]
\begin{center}
\includegraphics[scale=0.30]{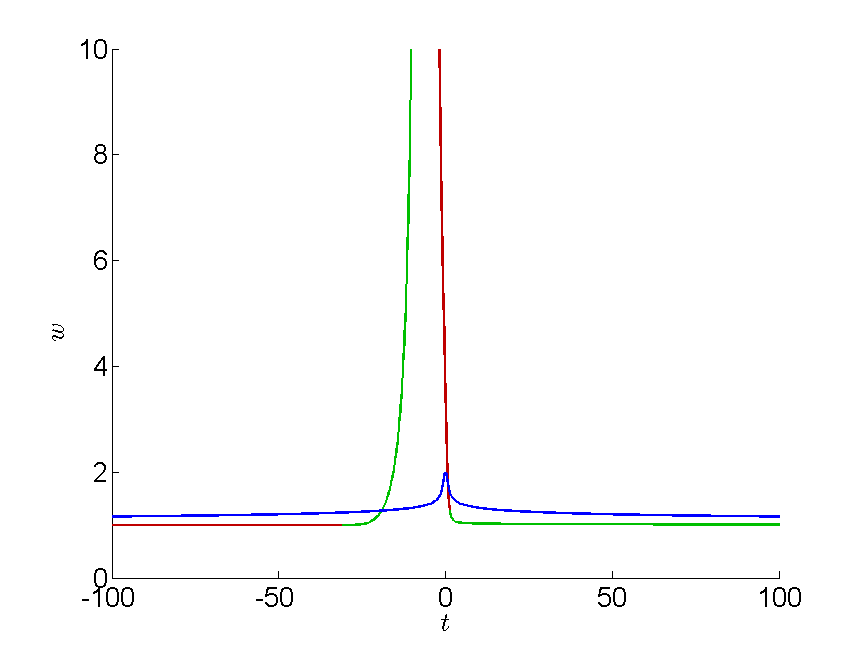}
\end{center}
\caption{Effective EoS parameter for the orbits shown in Figure \ref{cairebot}.}
\label{cairebotweff}
\end{figure}

Nevertheless, when a sufficiently low $V_0$ is used, orbits do not reach $\rho=0$ in a finite time and, hence, they have a single cycle, as we can see in Figures \ref{cai} and \ref{caiweff}.

\begin{figure}[H]
\begin{center}
\includegraphics[scale=0.32]{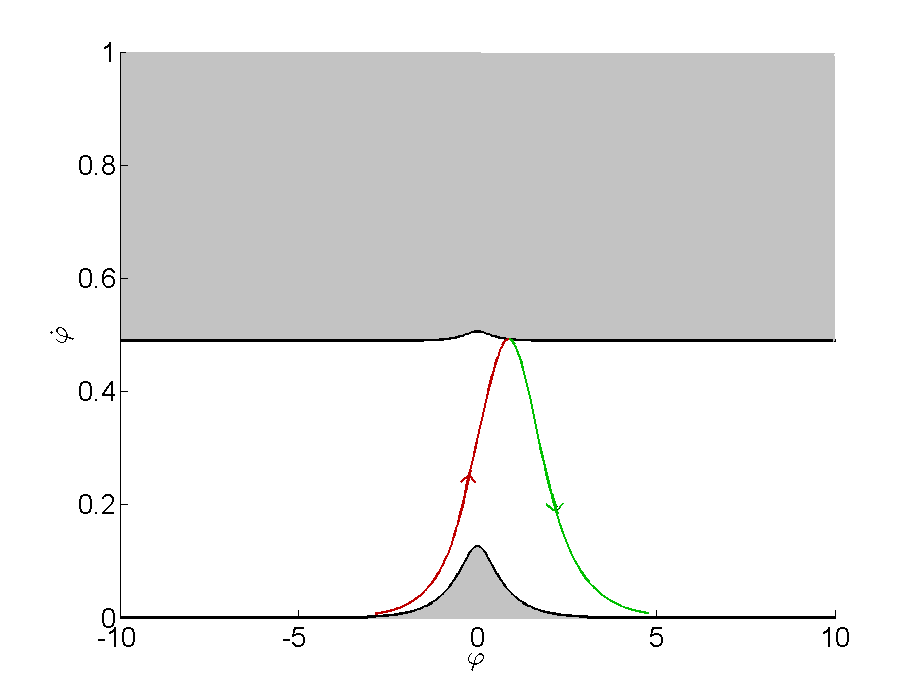}
\includegraphics[scale=0.34]{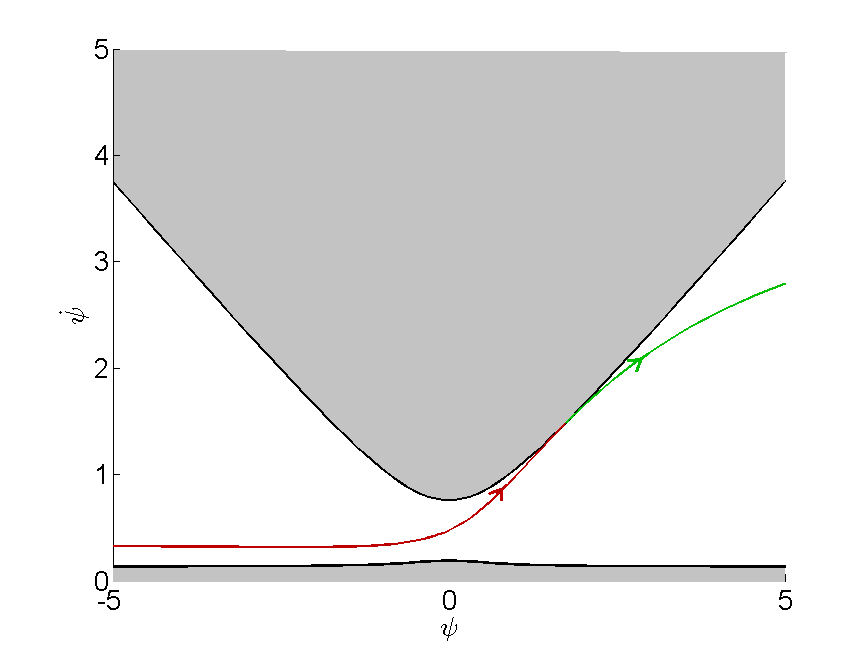}
\end{center}
\caption{Phase portrait in the $(\varphi,\dot{\varphi})$ plane (left) and in the $(\psi,\dot{\psi})$ plane for the LQC system
with potential \eqref{potcwe} for $w_0=2$ and $V_0=\frac{\rho_c(w_0-1)}{15}$.}
\label{cai}
\end{figure}

\begin{figure}[H]
\begin{center}
\includegraphics[scale=0.35]{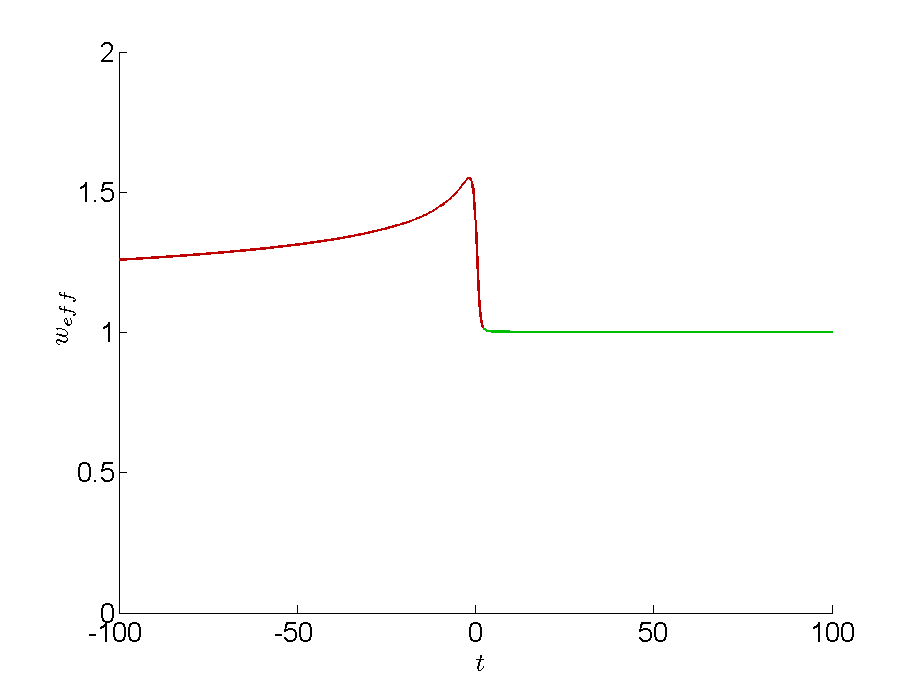}
\end{center}
\caption{Effective EoS parameter for the orbit shown in Figure \ref{cai}.}
\label{caiweff}
\end{figure}

 \end{enumerate}

 \section{The power spectrum in the matter-ekpyrotic bounce scenario}
In this section we will calculate explicitly the power spectrum of cosmological perturbations in the matter-ekpyrotic bounce scenario. In Loop Quantum Cosmology the 
corresponding Mukhanov-Sasaki equation 
\cite{sasaki}, 
in Fourier space,
is given by \cite{Cailleteau}
\begin{eqnarray}\label{ms}
 v''_k+c_s^2 k^2v_k-\frac{z''}{z}v_k=0,
\end{eqnarray}
where the velocity of sound is given by $c_s^2=1-\frac{2\rho}{\rho_c}$, $z=a\frac{\varphi'}{\mathcal H}$, being $'$ the derivative with respect the conformal time and
${\mathcal H}=\frac{a'}{a}=aH$ the conformal Hubble parameter.

 During the matter-domination period, { when $\rho\ll \rho_c \Longrightarrow c_s^2\cong 1$, i.e., far enough from the bounce in the contracting phase, the holonomy corrections could safely be disregarded and the equation (\ref{ms}) becomes the usual Mukhanov-Sasaki equation in GR
 \begin{eqnarray}\label{ms1}
 v''_k+ k^2v_k-\frac{z''}{z}v_k=0.
\end{eqnarray}

On the other hand, it is well-known that, in conformal time, when the universe is  matter dominated and the holonomy corrections can be
disregarded, the scale factor evolves as 
 follows \cite{wilson2}
\begin{eqnarray} \label{34} a(\eta)=a_E\left(\frac{\eta-\eta_1}{\eta_E-\eta_1}  \right)^2\Longrightarrow {\mathcal H}(\eta)=\frac{2}{\eta-\eta_1},
\end{eqnarray}
with $\eta_1=\eta_E-\frac{2}{{\mathcal H}_E}$, and $\eta_E$ is the transition time, where it is assumed that the phase transition occurs far from the bounce and, once again,
the GR is a good approximation.

Therefore, since at early times during the matter-domination, in the contracting phase, the energy density is smaller than the critical one, one can use the equation (\ref{ms1}) instead of (\ref{ms}), whose
 Bunch-Davies vacuum will be given by the well-known modes \cite{wilson2}
\begin{eqnarray}v_k(\eta)=-\sqrt{\frac{-\pi(\eta-\eta_1)}{4}}H_{3/2}^{(1)}[-k(\eta-\eta_1)]=\frac{e^{-ik(\eta-\eta_1)}}{\sqrt{2k}}\left(1-\frac{i}{k(\eta-\eta_1)}  \right).
\end{eqnarray}

The modes will eventually leave the Hubble radius, and thus,
when they are well outside of the Hubble radius, i.e., when they satisfy $|k(\eta-\eta_1)|\ll 1  \Longleftrightarrow \left| \frac{z''}{z} \right|\leq k^2$, 
\begin{eqnarray}v_k(\eta)\cong -\frac{4}{3\sqrt{2}}\frac{k^{3/2}}{{\mathcal H}^2(\eta)}-\frac{i}{2\sqrt{2}}\frac{{\mathcal H}(\eta)}{k^{3/2}}.
\end{eqnarray}

On the other hand, for these modes, up to their reentry in the Hubble radius in the expanding phase,  we can use the long-wavelenght approximation, and the Mukhanov-Sasaki equation  (\ref{ms}) becomes
\begin{eqnarray}
v''_k-\frac{z''}{z}v_k=0,\end{eqnarray}
whose solution is given by
\begin{eqnarray}v_k(\eta)\cong A_1(k)z(\eta)+A_2(k)z(\eta)\int_{-\infty}^{\eta}\frac{d\bar\eta}{z^2(\bar\eta)},
\end{eqnarray}
which is valid from the exit in the contracting phase to the reentry into the Hubble radius in the expanding one, and thus, also during the bounce.

\
To obtain the coefficients $A_1$ and $A_2$, we note that}
  in a matter dominated universe at early times, $z$ is given by $\sqrt{3}a$, because the holonomy corrections could be disregarded. A simple calculation leads to
\begin{eqnarray}v_k(\eta)\cong \sqrt{3}A_1(k)a_E\left(\frac{{\mathcal H}_E}{{\mathcal H}(\eta)} \right)^2-\frac{2}{3\sqrt{3}a_E}A_2(k)
\frac{{\mathcal H}(\eta)}{{\mathcal H}^2_E}.
\end{eqnarray}

Matching both expressions, we obtain the following values of $A_1$ and $A_2$
\begin{eqnarray}
 A_1(k)=  -\frac{4}{3\sqrt{6}}\frac{k^{3/2}}{a_E{\mathcal H}^2_E},\quad A_2(k)=\frac{3\sqrt{3}i}{4\sqrt{2}}\frac{a_E{\mathcal H}_E^2}{k^{3/2}},
\end{eqnarray}
and thus, the curvature fluctuation in co-moving coordinates is given by
\begin{eqnarray}\label{36}\zeta_k(\eta)\equiv\frac{v_k(\eta)}{z}\cong -\frac{4}{3\sqrt{6}}\frac{k^{3/2}}{a_E{\mathcal H}^2_E}+
\frac{3\sqrt{3}i}{4\sqrt{2}}\frac{a_E{\mathcal H}_E^2}{k^{3/2}}
\int_{-\infty}^{\eta}\frac{d\bar\eta}{z^2(\bar\eta)}\nonumber \\ 
\cong \frac{3\sqrt{3}i}{4\sqrt{2}}\frac{a_E{\mathcal H}_E^2}{k^{3/2}}
\int_{-\infty}^{\eta}\frac{d\bar\eta}{z^2(\bar\eta)}.
\end{eqnarray}

Note that, if $z$ is continuous, at the transition time $\eta=\eta_E$, the formula \eqref{36} holds for every time, and we can calculate the power spectrum as
\begin{eqnarray}\label{37}{\mathcal P}_{\zeta}(k)=\frac{k^3}{2\pi^2}|\zeta_k(\infty)|^2=\frac{27a_E^2{\mathcal H}_E^4}{64\pi^2}\left|
\int_{-\infty}^{\infty}\frac{d\eta}{z^2(\eta)}
\right|^2
\nonumber \\
=\frac{3{ H}_E^2}{64\pi^2}\left|
\int_{-\infty}^{\infty}\frac{a_Ez^2_E}{a(t)z^2(t)}H_Edt
\right|^2,
\end{eqnarray}
 where, when $\eta\rightarrow -\infty$ the asymptotic form of $a(\eta)$ is given by \eqref{34}, i.e., at very early times $a(t)\cong a_E\left(\frac{3}{2}H_E t\right)^{2/3}$.
This is essential in order to perform numerical calculations, because once an orbit has been chosen,  the value of $H(t)$ is given by  formula \eqref{25}, the 
value of $a(t)$ is obtained from the relation $\frac{\dot{a}(t)}{a(t)}=H(t)$, the asymptotic condition at early times $a(t)\cong a_E\left(\frac{3}{2}H_E t\right)^{2/3}$ and the value of 
$z(t)$ via the relation 
$z(t)=a(t)\frac{\dot\varphi(t)}{H(t)}$.

\

{
\begin{remark}
It is important to realize that formula  (\ref{36}) is only valid for the modes that leave the Hubble radius in the matter dominated epoch, i.e., when the holonomy corrections could be disregarded and the square of the velocity of sound is equal to $1$. However, in the so-called super-inflationary phase  ($ \rho_c/2 \leq \rho \leq \rho_c$),  the velocity
of sound becomes imaginary,  which could lead, during this stage, to a Jeans instability for ultra-violet
modes satisfying $k^2 |c_s^2| \gg \left| \frac{z''}{z}\right|$.

 Consequently, these sub-horizon growing modes could condensate, and after leaving the Hubble radius, they
could produce undesirable cosmological consequences.  Sometimes, in LQC, it is argued that  its  wavelength is too small  to be detected or directly this problem is not discussed, but really what happens is that, during 
this regime, the validity of the linear perturbations equations does not seem likely.
This is one of the reasons why a Teleparallel version of LQC has  been introduced
in \cite{haro}. This theory is based on the following two points: 1.- For
the FLRW geometry, the holonomy
corrected Friedmann equation (\ref{25}) could be obtained as a particular case of a
teleparallel $F(T)$ theory. 2.- The perturbation equations of $F(T)$ lead to a modified Mukhanov-Sasaki equation with a square of the velocity of sound always positive, and thus, avoiding the Jeans instability.

\end{remark}}

\

As an example, if one disregards the phase transition, i.e., if one only considers a matter domination during all the background evolution, 
and one uses the orbit given by the solution (\ref{28}) with $w_0=0$,
formula \eqref{37} leads to the result
${\mathcal P}_{\zeta}(k)= \frac{\pi^2}{9}\frac{\rho_c}{\rho_{Pl}}$ (see \cite{wilson} and \cite{haro} for a derivation of the result), where $\rho_{Pl}$ is the Planck's energy density,   which is incompatible with the experimental data  
${\mathcal P}_{\zeta}(k)\cong 2 \times 10^{-9}$ \cite{WMAP}, because in LQC the value of the 
critical density is given by $\rho_c\cong 258\cong 0.4 \rho_{Pl}$. This means that, to be viable,
the matter bounce scenario must be implemented, for example, with an ekpyrotic phase transition.

In this way we can introduce a transition.
However, when it is too  abrupt, formula \eqref{37} does not hold. To be more precise, we consider the simple example \cite{wilson2} where the scale factor is given, in 
terms of 
the cosmic
time, by
\begin{eqnarray}\label{cai-wilson}
 a(t)=\left\{\begin{array}{ccc}
  a_E\left(\frac{t-t_1}{t_E-t_1}\right)^{2/3}& \mbox{when} & t<t_E,\\
  \left(\frac{3\rho_c}{4}(1+w_0)^2t^2+1\right)^{\frac{1}{3(1+w_0)}}& \mbox{when} & t_E\leq t\leq t_D,\\
  a_D\left({3\rho_c}(t-t_2)^2+1\right)^{\frac{1}{6}}& \mbox{when} & t\geq t_D,
             \end{array}\right.
\end{eqnarray}
where $t_1\equiv t_E-\frac{2}{3H_E}$, being $H_E$ the value of the Hubble parameter at the time of the phase transition and $t_D\gtrsim 0$ is the time when the universe
enters in a deflationary or kination regime. So that the Hubble parameter be continuous at $t_D$ one has to choose
$$t_2=t_D-\frac{1}{6H_D}\left(1-\sqrt{1-\frac{12H_D^2}{\rho_c}}  \right)\cong t_D-\frac{H_D}{\rho_c}\cong \frac{(1-w_0)t_D}{2}\lesssim 0,
$$
where we have used that $H_D\cong \frac{\rho_c (1+w_0)t_D}{2}$.

In \cite{wilson2} the authors heuristically obtained that ${\mathcal P}_{\zeta}(k)\sim H_E^2$. Then, since the observational values of the power spectrum states
${\mathcal P}_{\zeta}(k)\cong 2\times 10^{-9}$, one will deduce that $H_E\sim 10^{-4}$.

To reproduce mathematically this result we consider,
for $\eta>\eta_E$, the solution in the long wave-lenght approximation
\begin{eqnarray}\label{40}\zeta_k(\eta)=B_1(k)+B_2(k)\int_{\eta_E}^{\eta}\frac{d\bar\eta}{z^2(\bar\eta)},\end{eqnarray}
then matching at $\eta=\eta_E$ one obtains
\begin{eqnarray}B_1(k)=A_1(k)+A_2(k)\int_{-\infty}^{\eta_E}\frac{d\bar\eta}{z^2(\bar\eta)},\quad B_2(k)=(1+w_0) A_2(k).\end{eqnarray}

That is,
\begin{eqnarray}B_1(k)=-\frac{4}{3\sqrt{6}}\frac{k^{3/2}}{a_E{\mathcal H}^2_E}-\frac{i}{2\sqrt{6}}\frac{{\mathcal H}_E}{a_Ek^{3/2}}
\cong -\frac{i}{2\sqrt{6}}\frac{{\mathcal H}_E}{a_Ek^{3/2}}= -\frac{i}{2\sqrt{6}}\frac{{ H}_E}{k^{3/2}},
\end{eqnarray}
because for modes well outside to the Hubble radius one has $k\ll |{\mathcal H}_E|$.

First we are interested in the evolution of  $\zeta_k(\eta)$ 
during the ekpyrotic regime.

For this reason  
we will calculate
\begin{eqnarray}B_2(k)\int_{\eta_E}^{\eta_D}\frac{d\bar{\eta}}{z^2(\bar\eta)}=\frac{\sqrt{3}ia_E^3{ H}_E^2}{4\sqrt{2}k^{3/2}}\int_{t_E}^{t_D}\frac{1-\frac{\rho}{\rho_c}}{a^3}dt,
\end{eqnarray}
where we have used that, during the ekpyrotic phase, 
\begin{eqnarray}
 z^2=a^2\frac{\dot{\varphi}^2}{H^2}=\frac{3(1+w_0)a^2}{1-\frac{\rho}{\rho_c}}.
\end{eqnarray}

Performing the change of variable $x=\frac{\rho}{\rho_c}\Longrightarrow dt=\frac{dx}{\sqrt{3\rho_c}(1+w_0)x^{\frac{3}{2}}\sqrt{1-x}}$ one obtains
\begin{eqnarray}B_2(k)\int_{\eta_E}^{\eta_D}\frac{d\bar\eta}{z^2(\bar\eta)}=
\frac{ia_E^3{ H}_E^2}{4\sqrt{2}(1+w_0)k^{3/2}\sqrt{\rho_c}}\left[\int^{1}_{\frac{\rho_E}{\rho_c}} {\sqrt{1-x}}\frac{1}{x^{\frac{(1+3w_0)}{2(1+w_0)}}}dx
\right. \nonumber \\ \left.
+\int_{\frac{\rho_D}{\rho_c}}^1{\sqrt{1-x}}\frac{1}{x^{\frac{(1+3w_0)}{2(1+w_0)}}}dx\right].
\end{eqnarray}

The integrals that appear in this expression are of the same order as
\begin{eqnarray}
\int^{1}_{\frac{\rho_E}{\rho_c}}{x^{-\frac{(1+3w_0)}{2(1+w_0)}}}dx\cong\frac{2(w_0+1)}{w_0-1}\left(\frac{\rho_c}{\rho_E}\right)^{\frac{w_0-1}{2(1+w_0)}},
\end{eqnarray}
\begin{eqnarray}
\int^{1}_{\frac{\rho_D}{\rho_c}}{x^{-\frac{(1+3w_0)}{2(1+w_0)}}}dx\cong 0
\end{eqnarray}
because in this model $t_D\gtrsim 0\Longrightarrow \rho_D\cong \rho_c$.

Finally, using that $a_E^3=\left(\frac{\rho_c}{\rho_E}\right)^{\frac{1}{1+w_0}}$, we will get
\begin{eqnarray}
 B_2(k)\int_{\eta_E}^{\eta_D}\frac{d\bar\eta}{z^2(\bar\eta)}\sim \frac{iH_E^2}{2\sqrt{2}(w_0-1)k^{3/2}\sqrt{\rho_c}}
 \left(\frac{\rho_c}{\rho_E}\right)^{\frac{1}{2}}=\frac{-iH_E}{2\sqrt{6}(w_0-1)k^{3/2}},
\end{eqnarray}
and, approximately, we have
\begin{eqnarray}
 \zeta_k(\eta_D)\cong \frac{-iw_0H_E}{2\sqrt{6}(w_0-1)k^{3/2}}.
\end{eqnarray}

During the deflationary phase, the curvature fluctuation evolves as
\begin{eqnarray}\zeta_k(\eta)=C_1(k)+C_2(k)\int_{\eta_D}^{\eta}\frac{d\bar\eta}{z^2(\bar\eta)}.\end{eqnarray}

Thus, performing the matching at $\eta=\eta_D$ one obtains
\begin{eqnarray}
 C_1(k)=\zeta_k(\eta_D)\cong \frac{-iw_0H_E}{2\sqrt{6}(w_0-1)k^{3/2}},\qquad C_2(k)=\frac{2}{1+w_0}B_2(k).
\end{eqnarray}

The deflationary regime ends when the universe becomes reheated and enters in a radiation dominated phase. For this reason we have to calculate
\begin{eqnarray}
 \zeta_k(\eta_R)=\zeta_k(\eta_D)+C_2(k)\int_{\eta_D}^{\eta_R}\frac{d\bar\eta}{z^2(\bar\eta)}\nonumber \\=
 \zeta_k(\eta_D)+\frac{i}{8\sqrt{2}}\left(\frac{a_E}{a_D} \right)^3 \frac{H_E^2}{k^{3/2}\sqrt{\rho_c}}
 \int_{\frac{\rho_R}{\rho_c}}^{\frac{\rho_D}{\rho_c}}\frac{\sqrt{1-x}}{x}dx.
\end{eqnarray}

Using the formulas $2.224$ and $2.225$ of \cite{gr}
$$
\int\frac{\sqrt{1-x}}{x}dx=
2\sqrt{1-x}+\ln\left|\frac{\sqrt{1-x}-1}{\sqrt{1-x}+1}  \right|,
$$
and the fact that $\frac{\rho_D}{\rho_c}\cong 1$ and $a_D\cong 1$, one obtains
\begin{eqnarray}
 \zeta_k(\eta_R)\cong
 \zeta_k(\eta_D)-\frac{i}{2\sqrt{2}}\left(\frac{\rho_c}{\rho_E} \right)^{\frac{1-w}{2(1+w)}}\frac{H_E}{k^{3/2}}
 \ln \left(\frac{\rho_c^{1/4}}{T_R}  \right),
\end{eqnarray}
where $T_R\sim \rho_R^{1/4}$ is the reheating temperature. Now, taking into account that in LQC $\rho_c\cong 252\sim 10^2$ and that the nucleosynthesis bounds impose
$4\times 10^{-22}\leq T_R\leq 4\times 10^{-10}$, one has
\begin{eqnarray}
 \zeta_k(\eta_R)\cong
 \zeta_k(\eta_D)-i\frac{H_E^{\frac{2w_0}{1+w_0}}10^{\frac{2}{1+w_0}}}{k^{3/2}}.
\end{eqnarray}

If we impose the first term on the right hand side to be greater than the second one, one will have the constraint
\begin{eqnarray}
 |H_E|\gg H_E^{\frac{2w_0}{1+w_0}}10^{\frac{2}{1+w_0}}\Longleftrightarrow |H_E|\ll 10^{-\frac{2}{w_0-1}}.
\end{eqnarray}

Now, with this assumption  one will have $\zeta_k(\eta_R)\cong
 \zeta_k(\eta_D)\cong \frac{-iw_0H_E}{2\sqrt{6}(w_0-1)k^{3/2}}.$

 When the universe enters in a radiation dominated phase one will have
 \begin{eqnarray}\zeta_k(\eta)=D_1(k)+D_2(k)\int_{\eta_R}^{\eta}\frac{d\bar\eta}{z^2(\bar\eta)},\end{eqnarray}
 where the holonomy corrections could be disregarded and then, $z\cong 2 a$ with $a(t)\cong a_R\left(\frac{t}{t_R}\right)^{1/2}$. Matching at $\eta=\eta_R$ one will get
 \begin{eqnarray}
  D_1(k)=\zeta_k(\eta_R),\qquad D_2(k)=\frac{2}{3}C_2(k).
 \end{eqnarray}

 A simple integration shows that, at very late times
 \begin{eqnarray}
  D_2(k)\int_{\eta_R}^{\eta}\frac{d\bar\eta}{z^2(\bar\eta)}=\frac{C_2(k)}{3a_R^3}\left[\frac{1}{2H_R}-\frac{t_R^{3/2}}{t}  \right]\cong \frac{C_2(k)}{6a_R^3H_R}.
 \end{eqnarray}

This last quantity is equal to
\begin{eqnarray}
 \frac{\sqrt{3}i}{4\sqrt{2}}\frac{H_E}{H_R}\left(\frac{a_E}{a_R} \right)^3 H_E\cong i\left(\frac{|H_E|}{10} \right)^{\frac{w_0-1}{1+w_0}}|H_E|,
\end{eqnarray}
where we have used that during the deflationary phase one has
\begin{eqnarray}
 a_R=a_D\left(\frac{\rho_R}{\rho_D} \right)^{-\frac{1}{6}}\cong \left(\frac{\rho_R}{\rho_c} \right)^{-\frac{1}{6}}.
\end{eqnarray}

Then, we conclude that, if $H_E$ satisfies the conditions
 $|H_E|\ll 10^{-\frac{2}{w_0-1}}$,
 one will have $\zeta_k(\infty)\cong \frac{-iw_0H_E}{2\sqrt{6}(w_0-1)k^{3/2}}$, which means that the power spectrum is given by
 \begin{eqnarray}
  {\mathcal P}_{\zeta}(k)\cong \frac{w_0^2H_E^2}{48\pi^2(1-w_0)^2},
 \end{eqnarray}
and thus, to match with observational data one has to choose 
{$H_E\sim \frac{1-w_0}{w_0}\times 10^{-3}$. Finally, the condition $|H_E|\ll 10^{-\frac{2}{w_0-1}}$ is fulfilled  for $w_0>{2}$.
}

\vspace{1cm}

\begin{remark}
As we have already explained, when one deals with a matter dominated Universe (without the ekpyrotic phase transition), the power spectrum of cosmological perturbations 
is of the order $\rho_c$, 
and since LQC provides a value   around Planck's density,
its value is in contradiction with the current observations because it is too high. This is another  reason to 
implement an ekpyrotic phase in the matter 
 bounce scenario.
\end{remark}

\section{The reheating process}

\begin{figure}[H]
\begin{center}
\includegraphics[scale=0.40]{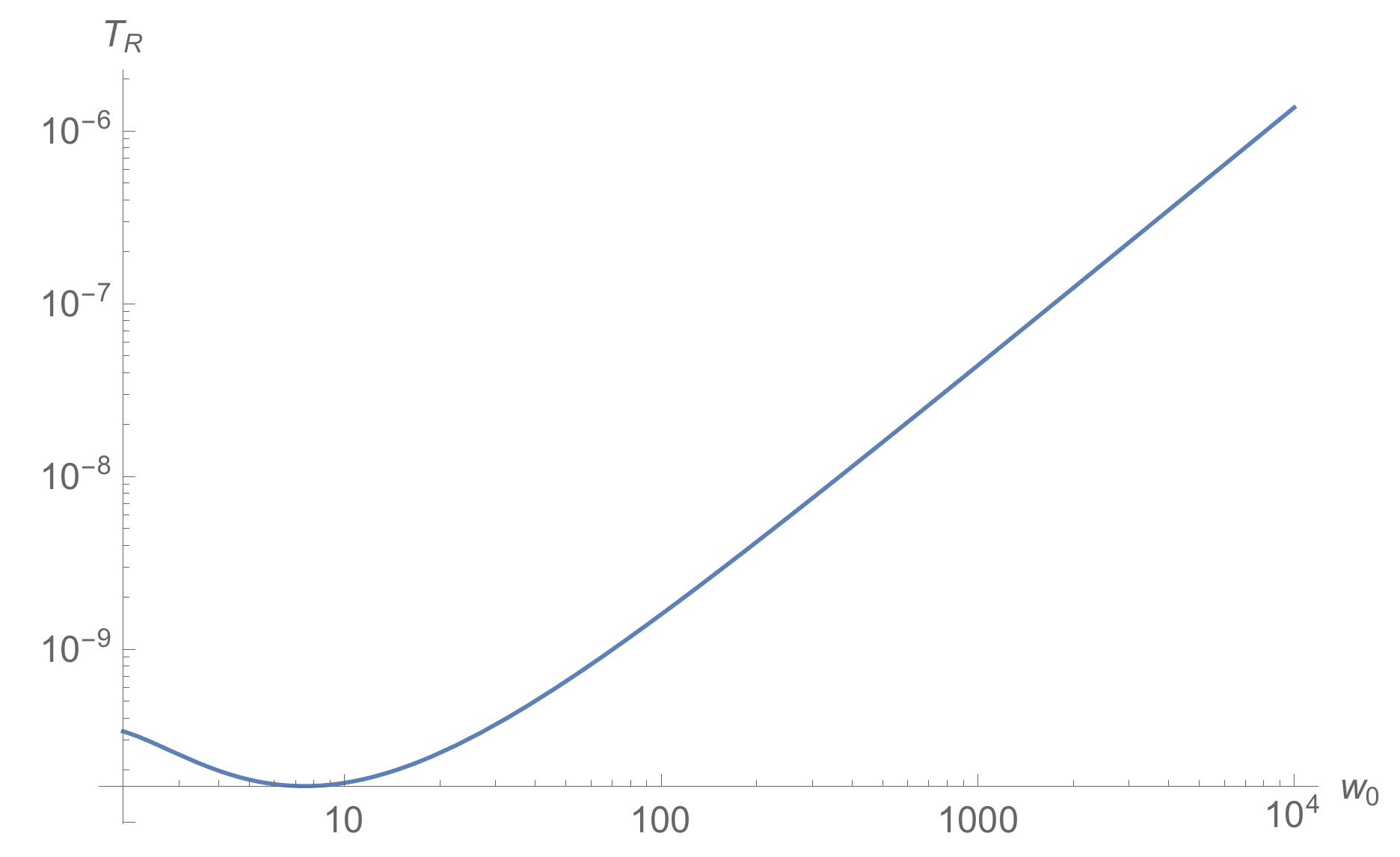}
\end{center}
\caption{The  reheating temperature as a function of  $w$ using formula (\ref{xxx}). }
\label{fig:Tvsw1}
\end{figure}

{
We will consider a heavy massive quantum field, namely $\chi$,  conformally coupled with gravity, and first of all we consider the  model (\ref{cai-wilson}) proposed by \cite{wilson2}. This model has been studied in \cite{he}, where it has been shown that  the energy density of the produced particles is given by
\begin{eqnarray}
\rho_{\chi}(t)=\frac{81 w_0^2 H_E^4}{4096 \pi}\left(\frac{a_E}{a(t)} \right)^4\sim 6\frac{(w_0-1)^4}{w_0^2}\times 10^{-15}\left(\frac{a_E}{a(t)} \right)^4 \equiv C\left(\frac{a_E}{a(t)} \right)^4 .
\end{eqnarray}

During the ekpyrotic phase, the background evolves as $\rho(t)=\rho_E\left(\frac{a_E}{a(t)} \right)^{3(1+w_0)}$, then, since $w_0>1$, during the contracting phase $\rho(t)$ increases faster than $\rho_{\chi}(t)$, and thus, the Universe continues being driven by the background. However, after the bounce, the Universe enters in a kination ($w_0=1$) regime and the background evolves as
$\rho(t)=\rho_c\left(\frac{a_c}{a(t)} \right)^{6}$, where 
\begin{eqnarray}\label{Y}
a_c=a_E \left(\frac{\rho_E}{\rho_c}\right)^{\frac{1}{3(1+w_0)}}\sim a_E \left(\frac{2\sqrt{3}(w_0-1)}{w_0} \right)^{\frac{2}{3(w_0+1)}}10^{-\frac{3}{1+w_0}}\equiv K a_E\end{eqnarray}
is the minimum value of the scale factor. Since in the expanding phase $\rho$ decreases as $a^{-6}$ and $\rho_{\chi}$ as $a^{-4}$, the energy density of the created particles will eventually start to dominate, and at that moment, namely $t_R$, when both energy densities will be of the same order, the Universe will become reheated with a temperature
$T_R\sim \rho^{1/4}(t_R)$.

To calculate this temperature we use the identity $\rho(t_R)\sim \rho_{\chi}(t_R)$ to obtain
\begin{eqnarray}
\frac{a_E}{a_R}\sim \sqrt{\frac{C}{\rho_c K^6}},
\end{eqnarray}
where $a_R=a(t_R)$. And, thus, the reheating temperature is given by
\begin{eqnarray}\label{xxx}
T_R\sim \frac{C^{3/4}}{\sqrt{\rho_c} K^3}
\end{eqnarray}

Taking into account, as we have shown in Section $4$, that for $w>2$ one has $|H_E|\sim \frac{w_0-1}{w_0}\times 10^{-3} $, one will obtain in Figure \ref{fig:Tvsw1} the  dependence of
$T_R$ as a function of $w_0$.

 }

\
\

Consequently, since nucleosynthesis bounds constrain the reheating temperature to be between $4\times 10^{-22}$  and $4\times 10^{-10}$,  from Figure \ref{fig:Tvsw1} we can see that the matter-ekpyrotic model 
only holds when $w$ is of the order $10$. However, the reheating temperature is very high and is in the borderline. Then, if one wants a lower reheating temperature, one has to consider a phase transition less abrupt than the one considered above. To do this, we will consider
 the model defined in next section,
where in the phase transition the second derivative
of the Hubble parameter is discontinuous.

Assuming that the mass of the quantum field is very high, for example is of the order of the reduced Planck mass,
we could approximate the vacuum modes by its  first order WKB solution, obtaining that the $\beta_k$ Bogoliubov coefficient is given by \cite{he,hap}
\begin{eqnarray}
 |\beta_k|^2\cong \frac{m^4a_E^{10}(\ddot{H}_E^+-\ddot{H}_E^-)^2}{256(k^2+m^2a_E^2)^5}\cong \frac{81m^4a_E^{10}w_0^2(1+w_0)^2 H_E^6}{1024(k^2+m^2a_E^2)^5 }.
\end{eqnarray}

The energy density of the produced particles is given by
\begin{eqnarray}
 \rho_{\chi}(t)=\frac{1}{2\pi^2 a^4(t)}\int_0^{\infty}k^2\sqrt{k^2+a^2(t)m^2}|\beta_k|^2dk\nonumber \\ \cong
 3 \left(\frac{3w_0(w_0+1)H_E^3}{128\pi m} \right)^2\left(\frac{a_E}{a(t)}\right)^4.
\end{eqnarray}

On the other hand, after the phase transition the energy density of the background evolves as $\rho(t)=\rho_E\left(\frac{a_E}{a(t)}\right)^{3(w_0+1)}$,
and after the bounce, since the universe enters in a kination domination, it evolves as 
$\rho(t)=\rho_c\left(\frac{a_c}{a(t)} \right)^{6}$, where  the value of the scale factor at the bouncing time is given by formula \eqref{Y}.

 In the contracting phase, the energy density is always dominant,
but in the expanding one it will eventually become subdominant. When both energy densities are of the same order the universe becomes reheated, this happens when
\begin{eqnarray}\label{72}
 \left(\frac{a_E}{a(t_R)} \right)^{2}\sim \frac{3}{\rho_c K^6}\left(\frac{3w_0(w_0+1)H_E^3}{128\pi m} \right)^2,
\end{eqnarray}
where the constant $K$ has been introduced in \eqref{Y}. Then,
the reheating temperature will be
\begin{eqnarray}\label{temperature}
 T_R\sim \rho^{1/4}(t_R)\sim \frac{D^{3/2}}{\sqrt{\rho_c} K^3} ,
\end{eqnarray}
where  $D\equiv \frac{(w_0-1)^2(w_0^2-1)}{m w_0^2} 10^{-11}$. And,
after reheating, the universe is dominated by a thermalized relativistic plasma and matches with the standard hot Friedmann universe.

\

\begin{figure}[H]
\begin{center}
\includegraphics[scale=0.40]{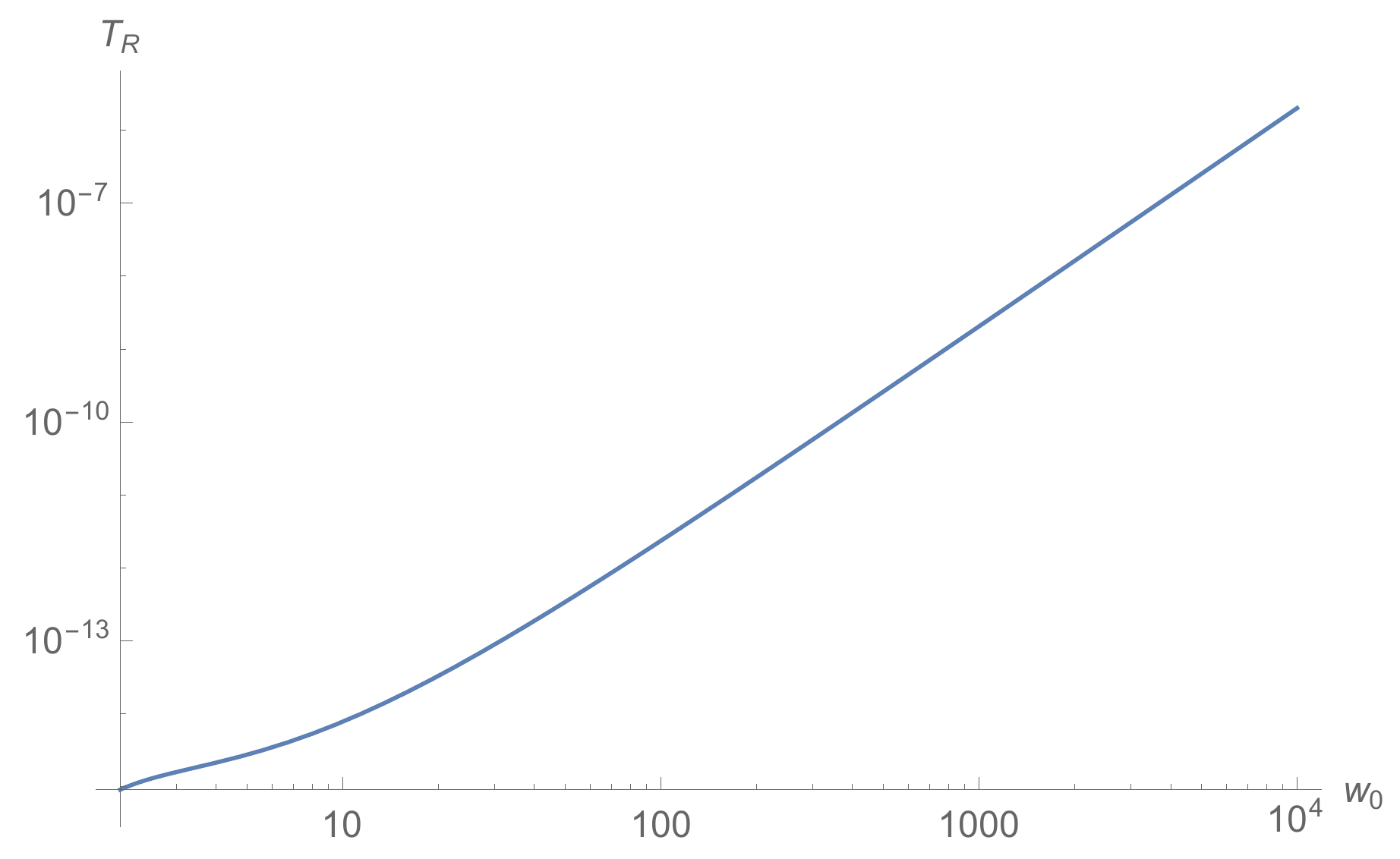}
\end{center}
\caption{The reheating temperature as a function of $w$ using formula \eqref{temperature}.}
\label{fig:Tvsw2}
\end{figure}

Finally, from formula \eqref{temperature}, 
if one takes, for instance, $m=M_{pl}=1$, 
one will obtain the picture drawn in Figure \ref{fig:Tvsw2}, which shows that the reheating temperatures  belong to the range between $4\times 10^{-22}$  and $4\times 10^{-10}$ for values
  of $w$ up to order $10^2$.

\

Two final remarks are  in order:

\begin{enumerate}
\item When one considers very light particles non-conformally coupled with gravity, the energy density of the produced particles is
$\rho_{\chi}(t)\sim 10^{-2} H_E^4\left( \frac{a_E}{a(t)} \right)^4$  \cite{ford,pv}, leading to the reheating temperature
\begin{eqnarray}\label{light}
 T_R\sim 
 3\times 10^{-10}\frac{(w_0-1)^3}{w_0^3K^3\sqrt{\rho_c}},
\end{eqnarray}
which, as we can see in Figure \ref{fig:Tvsw3}, is always greater that $2\times 10^{-11}$, i.e., the matter-ekpyrotic bounce scenario only supports the creation of massless particles 
when the reheating temperature of the universe is very high.

\begin{figure}[H]
\begin{center}
\includegraphics[scale=0.50]{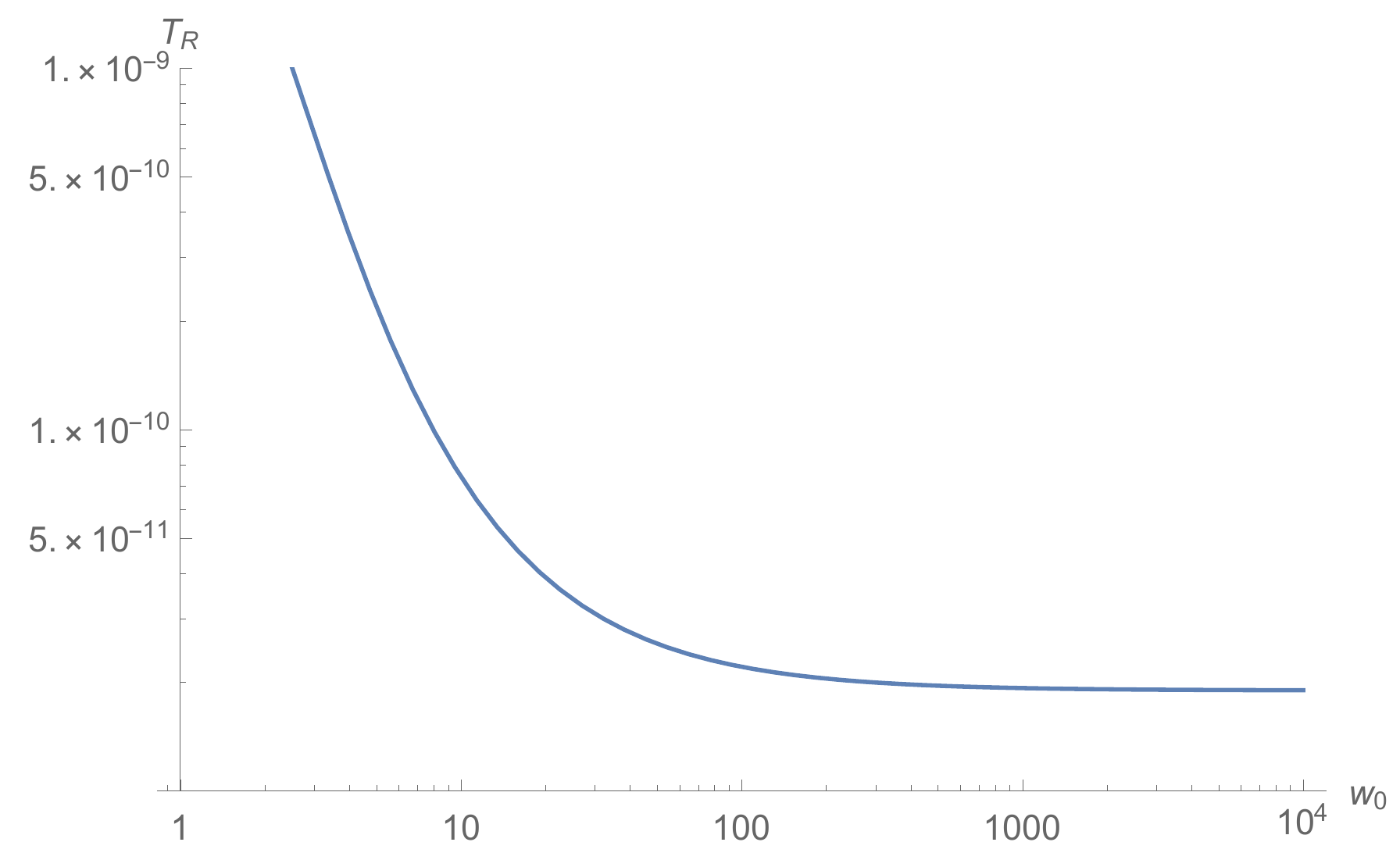}
\end{center}
\caption{The reheating temperature as a function of $w_0$ using the formula (\ref{light}).}
\label{fig:Tvsw3}
\end{figure}

\item The difference between the matter-ekpyrotic scenario and  the current models in quintessence inflation is that in the later one the phase transition occurs  after the end of inflation and, thus, the value of the Hubble parameter is of the order of $10^{-7}$ \cite{pv,hap},
 leading to a reheating temperature of the order   of $10^{-14}$, which is less than the one obtained in matter-ekpyrotic LQC scenario.
 \end{enumerate}

\section{Comparison with experimental data}
In the matter-ekpyrotic bounce scenario, for modes that leave the Hubble radius at very early times, when the universe is matter dominated, the spectral index and its running are given by
\cite{haro-cai, ewing}
\begin{eqnarray}
 n_s\cong 1+12w,\quad \alpha_s=\frac{n_s'}{(\ln{|\mathcal H}|)'}\cong -\frac{24\dot{w}}{H}(1-3w).
\end{eqnarray}

It is clear that a pure matter-ekpyrotic bouncing scenario leads to 
\begin{eqnarray}
 n_s=1\qquad \alpha_s=0.
\end{eqnarray}

Then, to match with the current observational data, one needs to improve this scenario introducing a quasi-matter domination period at very early times. 
To do this, first of all one has to consider the Raychaudhuri equation in LQC (see for instance \cite{singh1})
\begin{eqnarray}
\dot{H}=-\frac{P+\rho}{2}\left(1-\frac{2\rho}{\rho_c}  \right),\end{eqnarray}
which for the linear EoS $P=w_0\rho$ becomes
\begin{eqnarray}
\dot{H}=-\frac{(1+w_0)\rho}{2}\left(1-\frac{2\rho}{\rho_c}  \right).\end{eqnarray}

Then, in the ekpyrotic phase one has to simply choose $w_0>1$,  and for quasi-matter domination one could choose $w_0=-\epsilon$ (we have written $-\epsilon$ for convenience, but in principle $\epsilon$ can be positive or negative). However, at the phase transition, the derivative of the Hubble parameter is discontinuous and, as we have seen in the previous section, this kind of models could be ruled out by the nucleosynthesis bounds due to its high reheating temperature.  For this reason  we need to match these dynamics in a continuous way, and a simple way to do this is presented in the following model
\begin{eqnarray}
 \dot{H}=\left\{\begin{array}{ccc}
            -\frac{\rho}{2}\left(1-\epsilon+(w_0+\epsilon)\frac{\rho}{\rho_E}  \right)\left(1-\frac{2\rho}{\rho_c}  \right)& \mbox{when}& 0\leq \rho\leq \rho_E\\
             -\frac{(1+w_0)\rho}{2}\left(1-\frac{2\rho}{\rho_c}  \right)& \mbox{when}& \rho_E\leq \rho\leq \rho_c.
                \end{array}
\right.
\end{eqnarray}

Note that in the contracting phase, near $\rho_E$, since holonomy corrections could be disregarded, these dynamics could be written as follows
\begin{eqnarray}
 \dot{H}=\left\{\begin{array}{ccc}
            -\frac{3}{2}H^2\left(1-\epsilon+(w_0+\epsilon)\frac{H^2}{H_E^2}  \right)& \mbox{when}& 0\geq H\geq H_E\\
             -\frac{3(1+w_0)}{2}H^2& \mbox{when}&  H \lesssim H_E.
                \end{array}
\right.
\end{eqnarray}

Moreover, the effective Equation of State (EoS) parameter, namely $w=-1-\frac{2\dot{H}}{3H^2}$, is given by
\begin{eqnarray}
 w=\left\{\begin{array}{ccc}
                -\epsilon+(w_0+\epsilon)\frac{H^2}{H_E^2}& \mbox{when}& 0\geq H\geq H_E\\
                w_0& \mbox{when}&  H \lesssim H_E.
                \end{array}
\right.
\end{eqnarray}

Then, we can see that when $H\lesssim 0$ one has $w\cong -\epsilon$, that is, the universe is quasi-matter dominated.

\vspace{0.5cm}

On the other hand,
the potential that leads to these dynamics could be obtained approximately using the reconstruction method, as follows:
When holonomy corrections could be disregarded, Raychaudhuri equation becomes $\dot{H}=-\frac{P+\rho}{2}=-\frac{\dot{\varphi}^2}{2}$,
and then,
$\varphi=\int \sqrt{-2\dot{H}} dt=-\int \sqrt{\frac{-2}{\dot{H}}}dH$, obtaining for our model, when $0\geq H\geq H_E$
\begin{eqnarray}
\varphi=-\frac{2}{\sqrt{3(1-\epsilon)}}\arcsinh\left(\frac{H_E}{H}\sqrt{\frac{1-\epsilon}{w_0+\epsilon}} \right),
\end{eqnarray}
where we have used formula $2.266$ of \cite{gr}
$$\int\frac{dx}{x\sqrt{a+bx+cx^2}}= \frac{1}{\sqrt{a}}\arcsinh\left(\frac{2a+bx}{x\sqrt{4ac-b^2}} \right).
$$

Conversely,
\begin{eqnarray} H=-\frac{H_E\sqrt{1-\epsilon}}{\sinh\left({\frac{\sqrt{3(1-\epsilon))}}{2}\varphi} \right)\sqrt{w_0+\epsilon}}\cong 
\frac{2H_Ee^{\frac{\sqrt{3}}{2}\varphi}}{\sqrt{w_0}\left(1- e^{\sqrt{3}\varphi} \right)}.
\end{eqnarray}

Since, when holonomy corrections are disregarded, the potential is given by $V=3H^2+\dot{H}$, one has approximately for $0\geq H\geq H_E\Longleftrightarrow -\infty \leq\varphi \leq 
\varphi_E= -\frac{2}{\sqrt{3}}\ln\left(\frac{1}{\sqrt{w_0}}+\sqrt{\frac{1}{w_0}+1}\right)$
\begin{eqnarray}\label{quasimatter} 
V(\varphi)\cong \frac{6H_E^2e^{\sqrt{3}\varphi}}{w_0\left(1- e^{\sqrt{3}\varphi} \right)^2}\left(1-\frac{4e^{\sqrt{3}\varphi}}{\left(1- e^{\sqrt{3}\varphi} \right)^2}\right).
\end{eqnarray}

On the contrary, 
as we have seen in section $3$, in the ekpyrotic phase the potential is given by \cite{wilson}
\begin{eqnarray}
V(\varphi)=V_1\frac{e^{\sqrt{3(1+w_0)}\varphi}}{\left( 1+\frac{V_1}{2\rho_c}e^{\sqrt{3(1+w_0)}\varphi}\right)^2}, \quad \mbox{with}\quad w_0>1,
\end{eqnarray}
and,
since for the potential \eqref{quasimatter} one has $V(\varphi_E)\cong \frac{3}{2}H_E^2(1-w_0)$, assuming the continuity of the potential at the transition phase one gets
\begin{eqnarray}
V_1=\frac{4\rho_c^2(1-w_0)}{3H_E^2}e^{-\sqrt{3(1+w_0)}\varphi_E}\left( 1-\frac{3H_E^2}{2\rho_c} -\sqrt{ 1-\frac{3H_E^2}{\rho_c}} \right).
\end{eqnarray}

Therefore,
if one takes into account the approximation 
\begin{eqnarray*}1-x-\sqrt{1-2x}=1-x-\sqrt{(1-x)^2-x^2}
=(1-x)\left( 1-\sqrt{1-\frac{x^2}{(1-x)^2} } \right)\\
\cong \frac{x^2}{2(1-x)}\cong \frac{x^2}{2},
\end{eqnarray*}
 one will have
\begin{eqnarray} V_1\cong\frac{3}{2}H_E^2(1-w_0).
\end{eqnarray}

Summing up, the potential we propose in the matter-ekpyrotic bounce scenario is
\begin{eqnarray}
V(\varphi)=\left\{\begin{array}{ccc}
\frac{6H_E^2e^{\sqrt{3}\varphi}}{w_0\left(1- e^{\sqrt{3}\varphi} \right)^2}\left(1-\frac{4e^{\sqrt{3}\varphi}}{\left(1- e^{\sqrt{3}\varphi} \right)^2}\right)& \mbox{for} & \varphi\leq \varphi_E\\
V_1\frac{e^{\sqrt{3(1+w_0)}\varphi}}{\left( 1+\frac{V_1}{2\rho_c}e^{\sqrt{3(1+w_0)}\varphi}\right)^2} & \mbox{for} & \varphi\geq \varphi_E.
\end{array}\right.
\end{eqnarray}

Note that when  $\varphi\rightarrow -\infty$ the potential satisfies $V\propto e^{\sqrt{3}\varphi}$, that is, the universe is matter dominated at early 
times, and for $\varphi>\varphi_E$ the universe is in the ekpyrotic regime.

\vspace{1cm}

To perform the calculations, we choose, for example, as a pivot scale $k_*=a_0k_{phys}(t_0)$ where subindex $0$ means the current time, and
we choose, as usual, $k_{phys}(t_0) \sim 10^2H_0\sim 10^{-59}$ \cite{Ade}. To calculate the value of $k_*$ we need the current value of the 
scale factor which could be calculated as follows:

\

For the sake of simplicity we take $w_0=10\Longrightarrow |H_E|\sim 10^{-3}$. Then, from equation \eqref{72},
the value of the scale factor at the beginning of the radiation era will satisfy $a_R\sim 2\times10^9 a_E$.

 Now,   we use the relation  \cite{rg}
 \begin{eqnarray}
a_0\sim\frac{T_R}{T_0}a_R,
\end{eqnarray}
to obtain 
$a_0\sim 2\times 10^{30}a_E$, where we have used that the current temperature of the universe is $T_0\sim 8\times 10^{-32}$,
and that  for $w_0=1$ the reheating temperature is $T_R\sim 10^{-10}$. Consequently, 
$ k_*\sim  10^{-29}a_E$
and, thus since $|H_E|\sim 10^{-3}$, one concludes that $k_*\ll a_E|H_E|$.

\

On the other hand, 
let $t_*$ be the time when the pivot scale leaves the Hubble radius during the quasi-matter domination
in the contracting phase. Then one will have 
$a(t_*)|H(t_*)|=k_*\sim 10^{-29} a_E$, which means $|H(t_*)|\sim 10^{-29}\frac{a_E}{a(t_*)}\leq 10^{-29}$  because the pivot scale leaves the Hubble radius, in the contracting phase, before the phase transition. Then, we can conclude that $w_0\frac{H^2(t_*)}{H_E^2}\leq 10^{-51}$, and thus 
\begin{eqnarray}
 w\cong -\epsilon+ w_0\frac{H^2(t_*)}{H^2_E}\cong -\epsilon \Longrightarrow n_s\cong 1-12\epsilon,
\end{eqnarray}
and 
\begin{eqnarray}
 \alpha_s\cong -48w_0\frac{\dot{H}(t_*)}{H_E^2}=72w_0\frac{{H}^2(t_*)}{H_E^2}\geq 0.
\end{eqnarray}

Planck13 observational data at $1\sigma$ C.L. give the following results: $n_s=0.9583\pm 0.0081$ and $\alpha_s=-0.021\pm 0.012$. Then, to match our theoretical model,  with the $1$-marginalized domain for the spectral index at $2\sigma$ C.L.,
 one has to choose the  effective EoS parameter $w\cong \epsilon$ satisfying
\begin{eqnarray}
  0.0021\leq \epsilon\leq 0.0047.
\end{eqnarray}

On the other hand, since $72w_0\frac{H^2(t_*)}{H^2_E}\leq 10^{-49}$ we can see that the theoretical value of the running enters in  the $1$- marginalized domain for the running at $2\sigma$ C.L.

\begin{figure}[H]
\begin{center}
\includegraphics[scale=0.30]{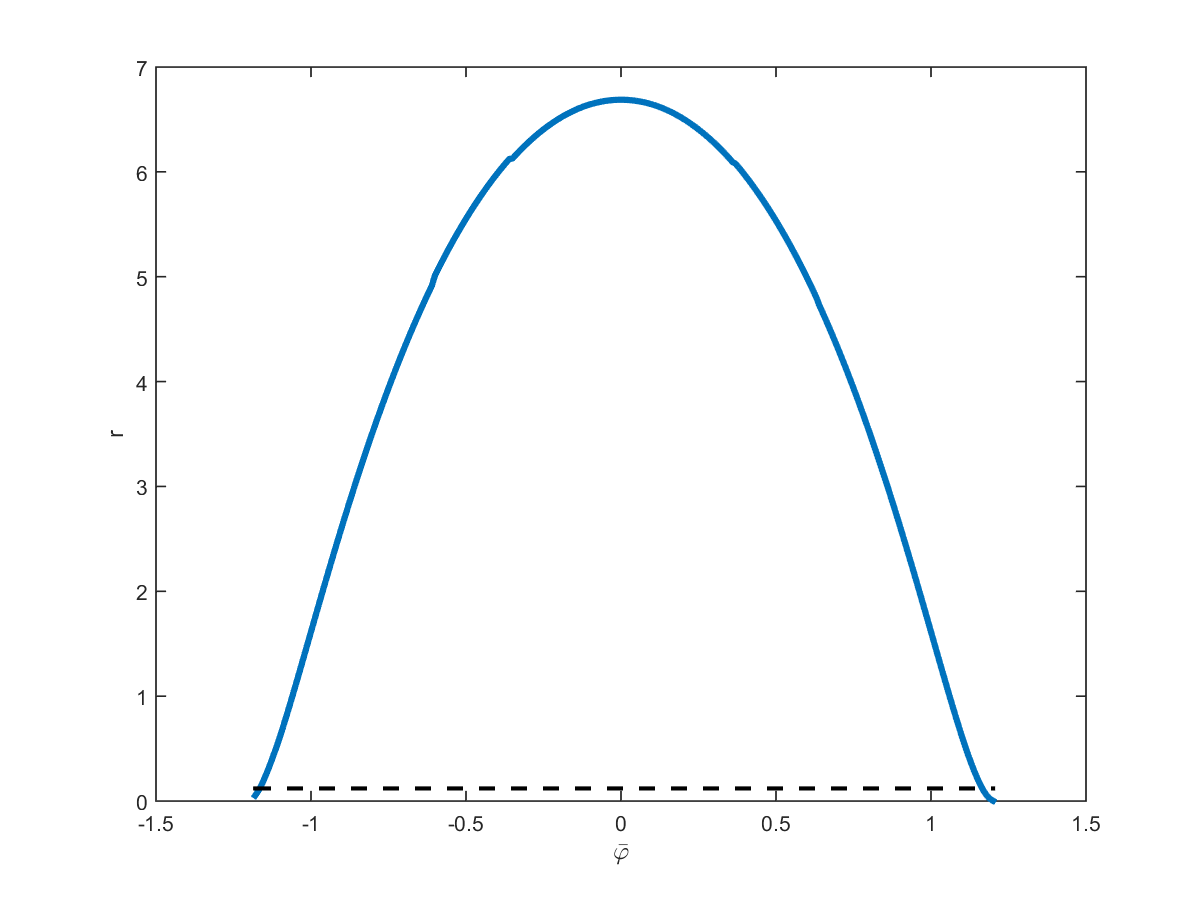}
\includegraphics[scale=0.30]{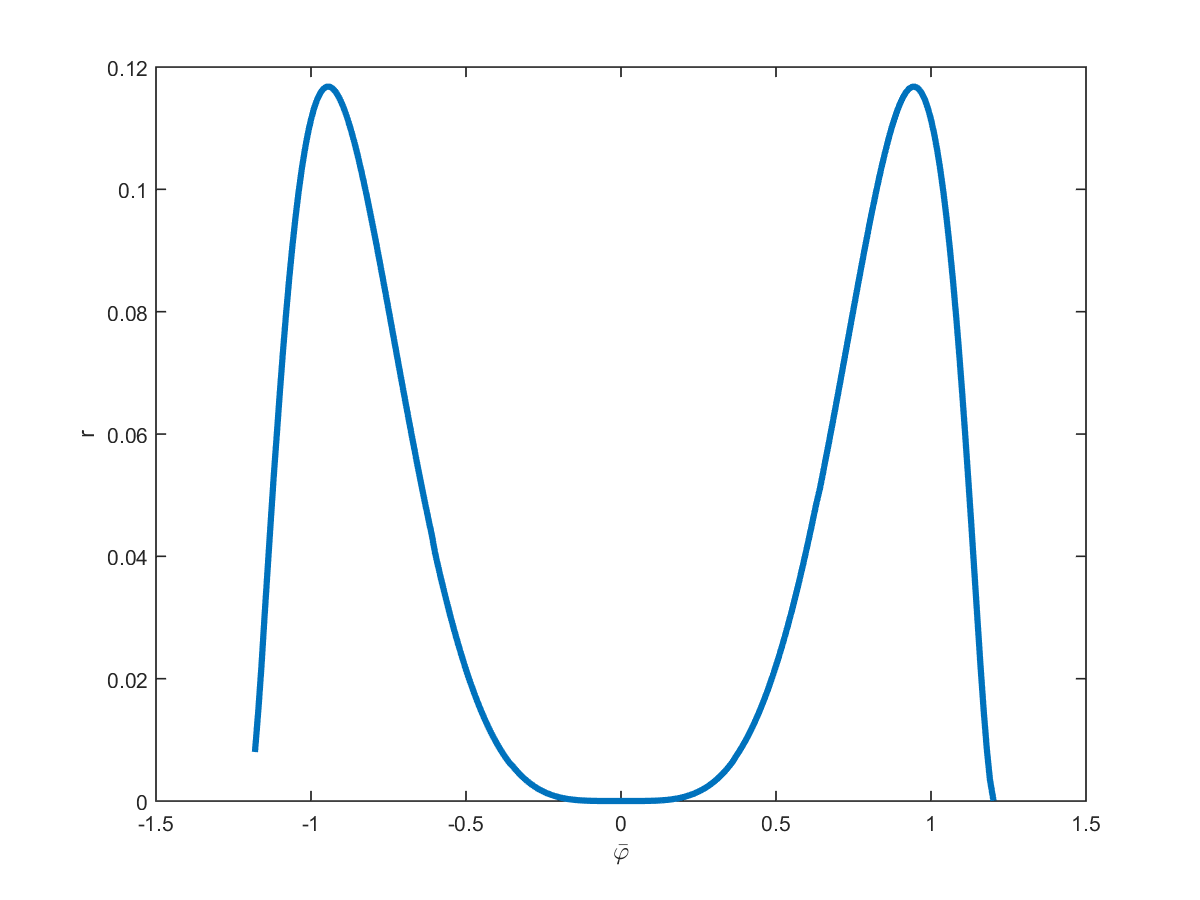}
\end{center}
\caption{{ The tensor to scalar ratio $r$ in a matter dominated scenario in LQC with teleparallel (left) or holonomy corrections (right), for a range of values of the bounce parameter $\varphi$. The potential of Section 4 of \cite{haro-amoros} and the corrected formula of \cite{haro_erratum} have been used.}}
\label{fig:rLQCmd}
\end{figure}

A final important remark is in order: As pointed out in \cite{wilson2}, in the matter-ekpyrotic scenario, dealing with the analytical solution, the power spectrum of tensor and scalar perturbations are of the same order. This means
that the ratio of tensor to scalar perturbations, namely $r$, does not satisfy the constraint $r\leq 0.12$ { provided by 
the joint analysis  data from BICEP2/Keck Array and Planck \cite{bicep2}}. There are several ways to address this problem, one of them is to consider at $\Lambda$CDM universe at early times \cite{cai1,cai2} because, due to the small velocity of sound, this increases the power spectrum of scalar perturbations and conserves the power spectrum of tensor ones, that is, it suppresses the tensor/scalar ratio. However, one has to realize that these calculations are performed for analytical solutions, but when one deals with a scalar field there are infinitely many solutions. { A quantitative study of the match of solutions to $r$
in a broad array of solutions has been done in a LQC matter-dominated scenario, 
both teleparallel and 
holonomy corrected versions, with $w_0=0$, in section 4.1 of \cite{haro-amoros}. 
Its results are summarized in Figure \ref{fig:rLQCmd}. The bound
$r\leq 0.12$ is satisfied in holonomy corrected LQC for all possible bounce values of the 
field $\varphi$, while in teleparallel corrected LQC the bound $r \le 0.12$ is satisfied
only at the tail cases $\varphi \le -1.16$ or $\varphi \ge 1.17$. In all cases
$r$ can go down to 0 in value. Whether the ekpyrotic regime in the early phase affects this match
of the theory with the expected value of $r$ is a difficult point that deserves future investigation.} 

\section{Conclusions}

In the present work we have shown, with all the mathematical details, that the matter-ekpyrotic bounce scenario in Loop Quantum Cosmology, whose background has a matter domination regime in the contracting phase at very early times followed by a phase transition to an ekpyrotic epoch up to the bounce where the universe enters in a kination phase, is a promising alternative to the inflationary paradigm, because it leads to theoretical values of some spectral parameters - the power spectrum of scalar perturbations, the spectral index and its running- that match at $2\sigma$ Confidence Level with the current observational data. Moreover, the phase transition, produced in the contracting phase,  is able to produce enough particles -very massive conformally coupled with gravity or massless non conformally coupled- to reheat the universe, in the expanding phase, at temperatures compatible with the bounds coming from nucleosynthesis, and thus, matching with the hot Friedmann universe.

{ On the other hand, the drawbacks affecting this proposed LQC matter-ekpyrotic 
bounce scenario include the uncertainty about the reliability of the  linear equations  of perturbations
for high $\rho$ in the sub-horizon $k$ regime due to the Jeans instability of these modes,
and the match of the predicted values of the tensor/ scalar perturbation ratio $r$ to its
observed bounds, which is still an open question.}

\vspace{1cm}
{\bf Acknowledgments.}
This investigation has been supported in part by MINECO (Spain), project MTM2014-52402-C3-1-P
and MTM2015-69135-P.

\end{document}